\documentclass[12pt]{article}

\usepackage{amsmath,amsthm,amssymb,latexsym,mathrsfs,longtable,epsfig,hhline,float,listings,color,graphicx,caption,subcaption}
\usepackage[margin=1in]{geometry}
\usepackage{bm}
\usepackage{tikz}
\usetikzlibrary{arrows,shapes}

\usepackage{lineno}

\newtheorem{theorem}{Theorem}[section]
\newtheorem{defn}{Definition}[section]

\usepackage[space]{grffile}

\graphicspath{{./Images/}}

\usepackage{color}
\newcommand{\red}[1]{\textcolor{red}{#1}}
\newcommand{\blue}[1]{\textcolor{blue}{#1}}

\def\Zh{\widehat Z}
\def\eps{{\varepsilon}}

\def\lth{\langle\!\langle}
\def\rth{\rangle\!\rangle}

\def\D{\mbox{${\rm D}$}}

\def\bw{{\bm \omega}}
\def\bq{{\bf q}}
\def\bk{{\bf k}}
\def\bO{{\bm \Omega}}
\def\bU{{\bm U}}
\def\be{{\bm \zeta}}
\def\bth{{\bm \theta}}
\def\bph{{\bm\phi}}
\def\bps{{\bm \psi}}

\def\bkap{{\bm \kappa}}

\numberwithin{equation}{section}

\begin{document}
\title{Genuine Nonlinearity and its Connection to the Modified Korteweg - de Vries Equation in Phase Dynamics}
\author{D.J. Ratliff\\
\emph{Department of Mathematics, Physics and Electrical Engineering,} \\\emph{Northumbria University,}\\ \emph{Newcastle upon Tyne, NE1 8ST, United Kingdom}}
\date{}
\maketitle

\begin{abstract}
The study of hyperbolic waves involves various notions which help characterise how these structures evolve. One important facet is the notion of \emph{genuine nonlinearity}, namely the ability for shocks and rarefactions to form instead of contact discontinuities. In the context of the Whitham Modulation equations, this paper demonstrate that a loss of genuine nonlinearity leads to the appearance of a dispersive set of dynamics in the form of the modified Korteweg de-Vries equation governing the evolution of the waves instead. Its form is universal in the sense that its coefficients can be written entirely using linear properties of the underlying waves such as the conservation laws and linear dispersion relation. This insight is applied to two systems of physical interest, one an optical model and the other a stratified hydrodynamics experiment, to demonstrate how it can be used to provide insight into how waves in these systems evolve when genuine nonlinearity is lost.
\end{abstract}

\section{Introduction}
The study of hydrodynamic systems remains at the heart of the study of nonlinear waves in modern physics. 
Ranging from the studies of fluids, optics, quantum mechanics and beyond~\cite{wlnlw}, they continue to prove their ability to be an accurate descriptor of observed phenomenon within such systems. 
Central to the study of this class of systems is are quantities known as characteristic speeds (or simply \emph{characteristics}) that reveal several properties about the nature of the system.
Primarily, they describe how information is transmitted in the problem but are also used to diagnose whether the underlying equations are hyperbolic or elliptic (real or complex characteristics respectively) which have implications for the stability of the states corresponding to such classifications. 
Another less frequent use of characteristics, which will be the key concept of this paper, is to diagnose genuine nonlinearity~\cite{l73}. This notion distinguishes whether nonlinear structures such as shocks and rarefactions can form. 
In cases where genuine nonlinearity is not operational, the system is said to be linearly degenerate and instead admits contact discontinuities.

An important hyperbolic system within the field of nonlinear waves, and one which will be the focus of the discussion of this paper, are the Whitham modulation equations (WMEs).
These govern the slow evolution of the wavenumbers and frequency of a given wave and thus determine its long-time evolution, which can be successfully applied to problems on wave stability~\cite{bnr14,bhj16}, the formation of dispersive shocks (\cite{eh16,ehs17} and references within) and localised structures~\cite{sh20}. However, the WMEs lack a regularisation mechanism such as dissipation or dispersion, but this can be remedied via a phase dynamical analysis to introduce dispersion into the system (see for example, refs.~\cite{bspmnw, br17,r19} and references therein). The form of the resulting dispersive equation has been shown to depend largely on properties of the characteristics admitted by the WMEs, with recent works highlighting significant changes in evolution in the neighbourhood of the elliptic-hyperbolic transition. The aim of this paper is to explore the result of the phase dynamics in light of a loss of genuine nonlinearity at a given state point (as opposed to a total linear degeneracy) to determine how this alters the evolution of the wave. The result of this reveals that the dispersive equation operational in such scenarios is the modified KdV (mKdV) equation:
\[
u_t\pm u^2 u_x+u_{xxx} = 0\,,
\]
where the function $u(x,t)$ is related to a perturbation of the wavenumber. The mKdV is a well-known nonlinear equation that arises across many fields, such as internal waves~\cite{cc99,dr78,gpt97,kb81}, plasma physics~\cite{ko69,ki74,gke92,nt94} and optics~\cite{a89,lgm14,tlm12}, and a key outcome of this paper is to unify the emergence of the mKdV in such environments by providing a universal derivation and form of the mKdV from a Lagrangian formalism, whose coefficients depend solely on properties of the wave from which they are derived.

The most straightforward derivation of the WMEs for a given single-phase wavetrain $\widehat{U}(kx+\omega t;k,\omega) \equiv \widehat{U}(\theta;k,\omega)$ involves an averaged Lagrangian approach \cite{wlnlw}, but these may also be obtained via formal asymptotics or by averaging the relevant conservation laws. In any of these cases, one arrives at the first order hydrodynamic system
\[
\begin{gathered}
k_T- \omega_X = 0\\[3mm]
A(k,\omega)_T+B(k,\omega)_X = 0\,,
\end{gathered}
\]
for local wavenumber $k(X,T)$ and frequency $\omega(X,T)$, slow variables $X = \eps x,\,T = \eps t$ with $\eps \ll 1$ and $A$, $B$ are the wave action and wave action flux.
The most well-understood version of these equations applies to single-phased wavetrains, as described above, but there have been generalisations to accommodate for both the cases of relative equilibrium, mean-flow effects and for arbitrarily many phases. For these cases, the WMEs generalise naturally to the vector-valued system
\begin{equation}
\begin{gathered}
\bk_T- \bw_X = {\bf 0}\\[3mm]
{\bf A}(\bk,\bw)_T+{\bf B}(\bk,\bw)_X = {\bf 0}\,,\\[3mm]
\bk,\,\bw,\,{\bf A},\,{\bf B} \in \mathbb{R}^N\,.
\end{gathered}
\end{equation}
Here, $\bk$ and $\bw$ are the vectors containing each slow wavenumber and frequency respectively and ${\bf A}$ and ${\bf B}$ are now the vector-valued wave action and wave action flux associated with each phase of the solution. Emerging from this system are up to $2N$ characteristics $c$ satisfying the zero determinant condition of a quadratic matrix pencil~\cite{br20}:
\begin{equation}
{\rm det}\big[c^2\D_\bw {\bf A}-c(\D_\bk{\bf A}+\D_\bw{\bf B})+\D_\bk{\bf B} \big] \equiv {\rm det}\big[{\bf E}(c)\big] = 0\,.
\end{equation}
The presence of a larger set of characteristics heralds an increasingly nontrivial set of ways in which these can interact and change the resulting dynamics for the system.

One shortcoming of the WMEs is that they lack a regularisation mechanism, such as dissipation or dispersion, which prevents gradient singularities and multivalued solutions from occurring. There has however been a recent series of approaches to remedy this issue via the use of phase dynamics. Inspired by the early works by the likes of Pomeau and Manneville~\cite{pm80}, Kuramoto~\cite{k84} and Doelman et al.~\cite{dsss}, Bridges et. al. adopted a similar modulational ansatz for use in Lagrangian systems to introduce dispersion into the modulation equations. Particularly, one very recent advancement utilises the characteristics to generate such dispersion in a general way. To do so, one constructs a guess at a new solution of the form
\begin{equation}\label{ansatz-intro}
\begin{split}
U = \widehat{U}(\theta+\eps \phi(X,T); k+\eps^2 \phi_X, \omega -\eps^2 c \phi_X+\eps^4 \phi_T)+W(\theta,X,T)\,,\\
 \quad X = \eps(x-ct)\,, \quad T = \eps^3 t\,, \quad \eps \ll 1\,.
\end{split}
\end{equation}
for phase $\theta = kx+\omega t$, phase perturbation $\phi$ and $c$ a real characteristic of the WMEs.
Substitution of the above into the Euler-Lagrange equations and a subsequent the asymptotic analysis leads to a dispersive set of dynamics emerging instead in the form of  the famous Korteweg - de Vries (KdV) equation,
\[
\big(\mathscr{B}_\omega+\mathscr{A}_k-2c\mathscr{A}_\omega\big)q_T+(\partial_k-c \partial_\omega)^2(\mathscr{B}-c\mathscr{A})qq_X+\mathscr{K}q_{XXX} = 0\,, \quad q = \phi_X\,,
\]
relying only on the fact that $c$ is real and thus the WMEs being hyperbolic~\cite{r19}. A remarkable feature of this analysis is that it demonstrates that the coefficients of this KdV are \emph{universal}, in the sense that they rely only on information regarding the conservation laws for the system rather than the particular form of the governing equations. Much like how the WMEs generalises to multiple phases, this approach and insight too extends naturally to waves with arbitrarily many phases

However, it can be shown that the form of the dispersive dynamics may alter dependent on the nature of the characteristics. For example, when characteristics coalesce at the elliptic-hyperbolic transition point it is instead the dynamics of the two-way Boussinesq which become operational~\cite{br17,br20}. Such a dynamical change heralds both quantitative and qualitative differences in how the system evolves - new solutions may arise, how they bifurcate may be altered and stability properties of solution families can change. This highlights that the properties of the characteristic can be used to diagnose which dispersive equation should be used to model the original wave's evolution as well as lending insight into how one expects such evolution to proceed.

It is in this spirit that the paper will proceed, with the main focus being on the connection between genuine nonlinearity and the resulting phase dynamics. The earliest work into this was undertaken by El et al.~\cite{ehs17}, who were able to demonstrate that a loss of genuine nonlinearity in a hydrodynamical system suggested the modified KdV (mKdV) equation should emerge with highly nontrivial consequences on the resulting dispersive shocks. The aim of this paper is to prove this connection generally for the WMEs, so that a loss of genuine nonlinearity for a given underlying wave signifies that an mKdV equation governs the dispersive dynamics of the wave quantities. We phrase this precisely as the following theorem:
\begin{theorem}\label{ther-paper}
Suppose a given Lagrangian admits an $N$-phased wavetrain solution. Then if the Lagrangian system is dispersive, the resulting Whitham modulation equations are hyperbolic, a chosen characteristic $c$ is simple and its field locally linearly degenerate for the wavetrain considered, the modified KdV equation,
\begin{equation}\label{mKdV-uni-multi}
\begin{split}
&\alpha U_T+\beta U^2U_X+\gamma U_{XXX} = 0\,,\\[4mm]
{\rm with} \qquad &\alpha = \be^T{\bf E}'(c)\be\,,\\[2mm]
&\beta  = -\frac{1}{2}\bigg[\be^T\bigg((\D_\bk-c\D_\bw)^2{\bf E}(c)\bigg)(\be,\be,\be)+3\be^T\bigg((\D_\bk-c\D_\bw){\bf E}(c)\bigg)(\be,{\bm \kappa})\bigg]\,,\\[2mm]
&\gamma = -\frac{1}{6}\sigma(0)''' \be^T{\bf E}'(c)\be\,,\\[2mm]
{\rm and} \qquad &{\bf E}(c)\be = {\bf 0} \,, \quad (\D_\bk-c\D_\bw){\bf E}(c)(\be,\be)+{\bf E}(c){\bm \kappa} = {\bf 0}\,,
\end{split}
\end{equation}
is an asymptotically valid reduction of its Euler-Lagrange equations in a frame moving with this characteristic speed.
\end{theorem}
The criterion outlined within the above statement are important in making sure that the coefficients of this reduction are nonzero. Hyperbolicity guarantees $c$ is real and the need for the simplicity of the characteristics ensures the coefficient of the time term, $\alpha$, doesn't vanish. The requirement for the system to be dispersive gives that $\gamma$ is nonzero except in special cases where the dispersion is weak. The loss of genuine nonlinearity is crucial to the presence of the cubic nonlinearity instead of the quadratic one the KdV possesses. The notion of local linear degeneracy, a relaxed form the classical linear degeneracy requirement, essentially states genuine nonlinearity is lost only at points and will be defined within the paper. This generically allows $\beta \neq 0$ and thus for nonlinearity to be retained within the phase dynamics in the form of the cubic term.

Once again the universality of the dispersive dynamics is apparent via the presence of the conservation laws (through ${\bf E}$ and its derivatives), however there is an additional universal feature to the above equation arising from the linear dispersion relation $\sigma$. The connection between the dispersive term in KdV-like models and the linear dispersion relation for the original system has long been known heuristically (for example, see~\cite{a94,g77,gos97}) but as of yet has not been rigorously proven for generic dispersive systems, and this paper provides such a proof. To do so, a Fourier-Bloch analysis  is presented inspired by preceding work~\cite{dsss}, and thus completely casts the coefficients of the resulting equation in terms of quantities obtainable from straightforward linear analyses. This lends a further strength to the analysis here - the nonlinear PDEs sought can be readily constructed from expressions one likely already possessed or those that are easily obtained, easing the access to information pertaining to the nonlinear evolution of the wave. 

 The essence of the proof of theorem \ref{ther-paper} is to adopt a rescaled and slightly modified version of the previously used ansatz (\ref{ansatz-intro}) so that all the terms in the mKdV asymptotically balance. The assumption of hyperbolicity, the use of a moving frame and the assumption of a loss of genuine nonlinearity then allow the analysis to proceed to the required order which the mKdV equation emerges at.

The appearance of the mKdV itself already sheds light on how a loss of genuine nonlinearity will affect the evolution of the underlying wave. Mainly it is the fact the mKdV admits a much larger solution set than the KdV equation, since the Muira transform connecting the two equations is not bijective~\cite{m68}, suggesting a more complex and richer evolution of the system. Further to the cnoidal and sech-based solitary wave solutions present in the KdV equation, dnoidal waves and front solutions connecting conjugate steady states also emerge, which grow as the square root of their speed rather than linearly with it as is the case for the KdV (see, for example, Grimshaw et al.~\cite{gpt97}). There are also breather and rational solutions which arise as solutions to the mKdV~\cite{aa18,zz14}. Stability properties of the two equations differs as well, for example periodic wave solutions of (\ref{mKdV-uni-multi}) can be modulationally unstable depending on the sign of the cubic nonlinearity, in stark contrast to the KdV where all such solutions are stable~\cite{bhj16}. With all of these factors considered it is clear that the transition in dynamics from the KdV equation to the mKdV equation via a loss of genuine nonlinearity presents a nontrivial set of changes to the overall evolution of the nonlinear wave.

Genuine nonlinearity plays an important role within the study of nonlinear waves across physics, albeit it is not always explicitly identified. Similarly, its loss and connection to the appearance of the mKdV within such systems remains widely unacknowledged. As part of the novelty of this work, it will be demonstrated how a loss of genuine nonlinearity can be identified in systems of physical interest, how the paper's theory can be used to construct the relevant mKdV in such scenarios and what the consequences of this might be for the original wave. In particular we focus on a higher order Nonlinear Schr\"odinger model utilised in the study of optical systems, which turns out to also provide information regarding the evolution of Stokes waves, in addition to a stratified shallow water system representing active experiments into internal solitary wave. This provides a template for how the theory of this paper may be utilised to understand the dynamics of nonlinear waves in situations where genuine nonlinearity is lost.

The paper proceeds in the following way. In \S\ref{sec:set-up}, the necessary abstract theory to undertake the phase dynamical approach is outlined and discussed. This includes a discussion of the wavetrain, the linearisation about it and the notion of genuine nonlinearity in the context of the WMEs. This is utilised in \S3 to prove theorem \ref{ther-paper} by constructing the relevant ansatz and undertaking a phase dynamical analysis. With the mKdV derived, we apply the theory to two examples in \S\ref{sec:examples}. The first is for a single phase wavetrain arising with an optical wave system, and the second appeals to an experimental set-up used to study internal waves in stratified fluids. Concluding remarks appear in \S\ref{sec:con}.

\section{Abstract Set-up and Linearisation Properties}\label{sec:set-up}

The starting point for the abstract theory discussed in this section is the multisymplectic form of the Lagrangian;
\[
\mathscr{L} = \iint \bigg[\frac{1}{2}\langle Z,{\bf M}Z_t+{\bf J}Z_x\rangle - S(Z) \bigg] \ dx\, dt\,,
\]
for skew symmetric matrices ${\bf M},\,{\bf J}$ and Hamiltonian function $S$. The procedure to transform a given Lagrangian into its respective multisymplectic form is a standard sequence of Legendre transforms~\cite{bspmnw}. The motivation for using this form is to provide a clear connection between the modulation analysis and the conservation laws, which enters through the matrices ${\bf M}$ and ${\bf J}$. The associated Euler-Lagrange equations in the multisymplectic formalism are then the variations of this Lagrangian:

\begin{equation}\label{ELE-MSF}
{\bf M}Z_t+{\bf J}Z_x = \D S(Z)\,.
\end{equation}
Throughout, the notation $\D$ will refer to the directional derivative
\[
\big[\D F(V)\big] R : = \lim_{\eps \rightarrow 0} \bigg( \frac{F(V+\eps R)-F(V)}{\eps}\bigg)\,,
\]
and the subscript, where present, will signify the argument being differentiated.

The theory of this paper proceeds under the assumption that the Euler-Lagrange equations (\ref{ELE-MSF}) has an $N$-phase wavetrain solution, where $N$ is a natural number, so explicitly we write this as
\[
\begin{split}
&Z = \Zh(\bth;\bk,\bw) \,, \\[3mm] \mbox{where} \quad 
\bth = 
\begin{pmatrix}
k_1 x+\omega_1 t +\theta_1^{(0)}\\
\vdots\\
k_N x+\omega_N t +\theta_N^{(0)}\\
\end{pmatrix}& = \begin{pmatrix}
\theta_1\\
\vdots\\
\theta_N
\end{pmatrix}\,, \quad
\bk = 
\begin{pmatrix}
k_1\\
\vdots\\
k_N
\end{pmatrix}\,, \quad
\bw = \begin{pmatrix}
\omega_1\\
\vdots\\
\omega_N
\end{pmatrix}\,.
\end{split}
\]
This wavetrain solution to (\ref{ELE-MSF}) satisfies the PDE
\begin{equation}\label{MJS-wavetrain}
\sum_{j=1}^N(\omega_j{\bf M}+k_j{\bf J}\big)\partial_{\theta_j}\Zh = \D S(\Zh)\,.
\end{equation}
Moreover, one may evaluate the Lagrangian along this wavetrain,
\[
\widehat{\mathscr{L}} = \iint \bigg[\frac{1}{2}\sum_{j=1}^N\langle \Zh,\omega_j{\bf M}\Zh_{\theta_j}+k_j{\bf J}\Zh_{\theta_j}\rangle - S(\Zh) \bigg] \ d\bth\,.
\]
Differentiating this with respect to the wavenumbers and frequencies leads to the wave action and wave action flux evaluated on the solution $\Zh$:
\begin{equation}\label{MJS-cons-defn}
\D_\bw \widehat{\mathscr{L}} = \frac{1}{2}
\begin{pmatrix}
\lth\Zh, {\bf M}\Zh_{\theta_1}\rth\\
\vdots\\
\lth\Zh, {\bf M}\Zh_{\theta_N}\rth\\
\end{pmatrix} = {\bf A}(\bk, \bw)\,, \quad \D_\bk \widehat{\mathscr{L}} = \frac{1}{2}
\begin{pmatrix}
\lth\Zh, {\bf J}\Zh_{\theta_1}\rth\\
\vdots\\
\lth\Zh, {\bf J}\Zh_{\theta_N}\rth\\
\end{pmatrix} = {\bf B}(\bk,\bw)\,.
\end{equation}
It is useful for the later analysis to take note of their derivatives with respect to wavenumber and frequency:
\[
\begin{split}
\D_\bw {\bf A}& = 
\begin{pmatrix}
\lth\Zh_{\omega_1}, {\bf M}\Zh_{\theta_1}\rth& \cdots & \lth\Zh_{\omega_N}, {\bf M}\Zh_{\theta_1}\rth\\
\vdots& \ddots& \vdots\\
\lth\Zh_{\omega_1}, {\bf M}\Zh_{\theta_N}\rth & \cdots& \lth\Zh_{\omega_N}, {\bf M}\Zh_{\theta_N}\rth
\end{pmatrix}\,, \\
\D_\bk {\bf A} & = 
\begin{pmatrix}
\lth\Zh_{k_1}, {\bf M}\Zh_{\theta_1}\rth& \cdots & \lth\Zh_{k_N}, {\bf M}\Zh_{\theta_1}\rth\\
\vdots& \ddots& \vdots\\
\lth\Zh_{k_1}, {\bf M}\Zh_{\theta_N}\rth & \cdots& \lth\Zh_{k_N}, {\bf M}\Zh_{\theta_N}\rth
\end{pmatrix}\,,\\
\D_\bw {\bf B}& = 
\begin{pmatrix}
\lth\Zh_{\omega_1}, {\bf J}\Zh_{\theta_1}\rth& \cdots & \lth\Zh_{\omega_N}, {\bf J}\Zh_{\theta_1}\rth\\
\vdots& \ddots& \vdots\\
\lth\Zh_{\omega_1}, {\bf J}\Zh_{\theta_N}\rth & \cdots& \lth\Zh_{\omega_N}, {\bf J}\Zh_{\theta_N}\rth
\end{pmatrix} = \D_\bk{\bf A}^T\,, \\
 \D_\bk {\bf B} & = 
\begin{pmatrix}
\lth\Zh_{k_1}, {\bf J}\Zh_{\theta_1}\rth& \cdots & \lth\Zh_{k_N}, {\bf J}\Zh_{\theta_1}\rth\\
\vdots& \ddots& \vdots\\
\lth\Zh_{k_1}, {\bf J}\Zh_{\theta_N}\rth & \cdots& \lth\Zh_{k_N}, {\bf J}\Zh_{\theta_N}\rth
\end{pmatrix}\,,\\
\end{split}
\]
as these matrices will arise within the definition of the characteristics, as well as a central feature of the phase dynamical analysis.

It is from these definitions for the conservation law components that we are able to discuss characteristics, which are the fundamental construct of the majority of this paper. The WMEs for $N$-phased wavetrain can be written as
\begin{equation}\label{WME-multi}
\begin{gathered}
{\bf K}_T-{\bm \Omega}_X = {\bf 0}\,,\\
{\bf A}({\bf K},{\bm \Omega})_T+{\bf B}({\bf K},{\bm \Omega})_X = {\bf 0}\,,
\end{gathered}
\end{equation}
for local vector-valued wavenumber ${\bf K}(X,T)$ and local frequency ${\bm \Omega}(X,T)$. The characteristics for this system about a fixed wavenumber and frequency $\bk,\,\bw$ can be found using the
 normal mode approach $({\bf K},\,{\bm \Omega}) = (\bk,\bw)+\delta(\hat{\bf K},\,\hat{\bm \Omega})\exp(i(X-cT))$, which to order $\delta$ gives the quadratic matrix pencil
\begin{equation}\label{E-defn}
c^2\D_\bw{\bf A}-c(\D_\bk{\bf A}+\D_\bw{\bf B})+\D_\bk{\bf B} \equiv {\bf E}(c)\,.
\end{equation}
The roots of its determinant, 
\begin{equation}\label{char-defn}
\Delta(c) \equiv {\rm det}\big[{\bf E}(c)\big] = 0\,,
\end{equation}
define the characteristics for the system. For these choices of $c$, we can define the eigenvector $\be$ satisfying
\begin{equation}\label{eig-defn}
{\bf E}(c)\be = {\bf 0}
\end{equation}
Throughout the paper, we will be assuming that the characteristic chosen is simple so that the kernel of ${\bf E}(c)$ is one dimensional and no other eigenvectors need be considered. As a consequence, it means that
\[
\Delta'(c) = {\rm tr}({\rm adj}({\bf E}){\bf E}') \propto \be^T(2c\D_\bw {\bf A}-\D_\bk{\bf A}-\D_\bw{\bf B})\be \neq 0\,,
\]
where the prime denotes differentiation with respect to $c$, tr denotes the trace of the matrix, $T$ the matrix transpose and adj denotes the matrix adjugate.
Interestingly, it is the final expression above that emerges as the coefficient of the time term in the mKdV, emphasising why the assumption on the characteristic's simplicity is necessary for its derivation.

\subsection{Linearisation Properties and Fourier-Bloch Analysis}\label{sec:JC-theory}
For the analysis leading to the modified KdV we must consider the linearisation of the system (\ref{MJS-wavetrain}) and its properties. Therefore, define the linear operator
\[
{\bf L}V = \big[\D^2S(\Zh)-\sum_{j=1}^N(\omega_j{\bf M}+k_j{\bf J})\partial_{\theta_j}\big]V\,.
\]
We can show that by taking $\theta_j$ derivatives of (\ref{MJS-wavetrain}) that
\begin{equation}\label{theta-deriv}
{\bf L}\Zh_{\theta_j} = 0\,,
\end{equation}
so that it is clear that $\Zh_{\theta_j}$ lies in the kernel of ${\bf L}$. We make the assumption that the kernel of ${\bf L}$ is no larger than the span of these elements, so that ker$({\bf L}) = {\rm span}\big\lbrace\Zh_{\theta_j}:j\in \lbrace 1,\ldots,N\rbrace \big \rbrace$. As such, the condition that a given expression lies in the range of ${\bf L}$ can then be formulated as
\begin{equation}\label{solvability}
G \mbox{ lies in the range of} \ {\bf L} \ {\rm if} \ \lth \Zh_{\theta_j},G\rth = 0 \ {\rm for} \ j = 1,\ldots,N\,,
\end{equation}
where $\lth \bullet,\bullet\rth$ is a suitable inner product for the problem. For multiply $2 \pi$-periodic waves, the natural choice is the averaging inner product over each phase:
\[
\lth U, V \rth = \frac{1}{(2 \pi)^N}\int_{[0,2 \pi)^N} \langle U,V \rangle \ d \bth\,.
\]
It is also necessarily for the analysis leading to the modified KdV equation to consider derivatives of (\ref{MJS-wavetrain}) with respect to the wavenumber and frequency. Doing so gives
\[
{\bf L}\Zh_{k_j} = {\bf J}\Zh_{\theta_j}\,, \quad {\bf L}\Zh_{\omega_j} = {\bf M}\Zh_{\theta_j}\,,
\]
which may be combined into the single expression
\begin{equation}\label{k-w-deriv}
{\bf L}(\Zh_{k_j}-c \Zh_{\omega_j}) = ({\bf J}-c{\bf M})\Zh_{\theta_j}\,,
\end{equation}
where $c$ is a constant to be determined shortly.
This suggests a Jordan chain structure is present, as is discussed in Bridges and Ratliff~\cite{br17}. The details are briefly recounted here, but for further details the reader is referred to this work instead. Two chains emerge, one of length four and one of length two, and it is the former we are concerned with. It takes the form
\[
{\bf L}{\bf v}_1 = 0\,, \quad {\bf L}{\bf v}_{j+1} = {\bf K}{\bf v}_j\,, \quad {\bf K} = {\bf J}-c{\bf M}\,,
\]
with the first two elements,
\begin{equation}\label{JC-elements}
{\bf v}_1 = \sum_{j=1}^N \zeta_j\Zh_{\theta_j}\,, \quad {\bf v}_2 = \sum_{j=1}^N \zeta_j (\Zh_{k_j}-c \Zh_{\omega_j})\,,
\end{equation}
for constants $\zeta_j$, following from (\ref{theta-deriv}) and (\ref{k-w-deriv}) respectively. The third element, defined by
\[
{\bf L}{\bf v}_3 = {\bf K} \sum_{j=1}^N \zeta_j (\Zh_{k_j}-c \Zh_{\omega_j})
\]
may be found providing the right hand side is in the range of ${\bf L}$. Assessing this using (\ref{solvability}) leads to the vector system
\[
\begin{pmatrix}
\lth \Zh_{\theta_1}, {\bf K}\sum_{j=1}^N \zeta_j (\Zh_{k_j}-c\Zh_{\omega_j})\rth\\
\vdots\\
\lth \Zh_{\theta_N},{\bf K}\sum_{j=1}^N \zeta_j (\Zh_{k_j}-c\Zh_{\omega_j})\rth
\end{pmatrix} = -\big(c^2\D_\bw{\bf A}-c(\D_\bk{\bf A}+\D_\bw{\bf B})+\D_\bk{\bf B}\big)\be \equiv -{\bf E}(c)\be= {\bf 0}\,.
\]
This is satisfied providing that $c$ is a characteristic of the Whitham modulation equations associated with the wavetrain $\Zh$ and the vector $\be$ is the eigenvector associated with the zero eigenvalue of ${\bf E}$ defined in (\ref{eig-defn}). The zero eigenvalue of ${\bf L}$ is even, and so the existence of ${\bf v}_3$ automatically guarantees the existence of ${\bf v}_4$ with
\[
{\bf L}{\bf v}_4 = {\bf K}{\bf v}_3\,.
\]
The length of the chain is precisely four when the right hand side of the expression for the next element of the chain,
\[
{\bf L}{\bf v}_5 = {\bf K}{\bf v}_4\,,
\]
does not lie in the range of ${\bf L}$. Within the analysis contained within this paper, we make this assumption as in practice it is the most generic case, only failing in special cases where dispersion is sufficiently weak. By (\ref{solvability}), this is precisely when
\[
\begin{pmatrix}
\lth \Zh_{\theta_1},{\bf K}{\bf v}_4\rth\\
\vdots\\
\lth \Zh_{\theta_N},{\bf K}{\bf v}_4\rth
\end{pmatrix} \equiv -{\bf T} \neq {\bf 0}\,.
\]

Surprisingly, the termination of this Jordan chain may be related to the linear dispersion relation obtained about the solution $\Zh$. The details of how this connection are a novel aspect of this paper and will be discussed below, and the result is simply that
\[
{\bf T} =  \frac{1}{6}\sigma'''(0)(2c\D_\bw{\bf A}-\D_\bk{\bf A}-\D_\bw{\bf B})\be  = \frac{1}{6}\sigma(0)'''{\bf E}'(c) \be
\]
where $\sigma(\nu)$ is the linear dispersion relation about the solution $\Zh$. This will go on to form the coefficient of the dispersive term in the mKdV derived in this paper. In this light the connection between this coefficient and the linear dispersion relation is natural, as the linear dispersion relation for the resulting mKdV equation must match the long wave expansion of the linear dispersion relation of the system for which it is derived.

The starting point to establish this connection is to introduce the Bloch ansatz
\[
\chi = Z(\bth,\nu)e^{i (\nu x-\sigma(\nu)t)}\,,
\]
where $\sigma(\nu)$ denotes a continuous set of eigenvalue curves which may be indexed~\cite{dsss}, however we will restrict ourselves to but a single one of these as will be made clear shortly. This Bloch form suggests the definition of the Bloch operator
\begin{equation}\label{Bloch_op}
{\bf L}_B = \D^2S(\Zh)-\sum_{j=1}^N\big({\bf M}(\omega_j\partial_{\theta_j}-i \sigma)+{\bf J}(k_j\partial_{\theta_j}+i \nu\big)\,,
\end{equation}
which we assume has $Z$ as a kernel element, so that
\begin{equation}\label{Bloch_lin}
{\bf L}_B Z = 0\,.
\end{equation}
The adjoint operator of this under the averaging inner product is simply it's complex conjugate.
When $\nu$ is taken to be zero, we have the eigenvalue problem
\[
({\bf L}+i\sigma(0) {\bf M}) Z = 0\,,
\]
Thus, $\sigma(0)$ needs to be an eigenvalue of the original linear operator. We choose this, quite naturally, to be the zero eigenvalue so that the discussion corresponds to the linear theory about $\Zh$. In doing so, it becomes clear that $\sigma(\nu)$ is the linear dispersion relation about $\Zh$ and we can write $Z(\bth,0)$ as a linear combination of the kernel of ${\bf L}$,
\begin{equation}\label{Z_nu_form}
Z(\bth,0) = \sum_{j = 1}^N\alpha_j \Zh_{\theta_j}\,.
\end{equation}
This can be ensured by assuming that ${\bf L}$ only has a simple zero eigenvalue.
Much of the discussion revolves around taking $\nu$ derivatives of the Bloch linearisation and evaluating these at $\nu = 0$. 
This is to essentially consider a long wave expansion of $\sigma$, and we will show that the Jordan chain structure discussed above arises naturally from doing this.
 Thus, differentiate (\ref{Bloch_lin}) with respect to $\nu$ four times and set $\nu = 0$:
\begin{subequations}
\begin{align}
{\bf L}\Zh_\nu(\bth,0)& = i \big({\bf J}-\sigma'(0){\bf M}\big)Z(\bth,0)\,,\label{Z_nu_eqtn}\\
{\bf L}Z(\bth,0)_{\nu \nu}& = 2i \big({\bf J}-\sigma'(0){\bf M}\big)Z(\bth,0)_\nu-i\sigma''(0){\bf M}Z\,,\label{Z_nu_2_eqtn}\\
{\bf L}Z(\bth,0)_{\nu \nu \nu}& = 3i \big({\bf J}-\sigma'(0){\bf M}\big)Z(\bth,0)_{\nu \nu}-3i\sigma''(0){\bf M}Z_\nu-i\sigma'''(0){\bf M}Z\,,\label{Z_nu_3_eqtn}\\
\begin{split}
{\bf L}Z(\bth,0)_{\nu \nu \nu \nu}& = -4i \big({\bf J}-\sigma'(0){\bf M}\big)Z(\bth,0)_{\nu \nu \nu}-6i\sigma''(0){\bf M}Z_{\nu \nu}\\
&\hspace{49mm}-4i\sigma'''(0){\bf M}Z_{\nu}-i\sigma''''(0){\bf M}Z\,.\label{Z_nu_4_eqtn}
\end{split}
\end{align}
\end{subequations}
The first equation, once (\ref{Z_nu_form}) is used, resembles the twisted Jordan chain result (\ref{k-w-deriv}).It follows that
\[
\Zh_\nu(\bth,0) = i\sum_{j = 1}^N\alpha_j (\Zh_{k_j}-\sigma'(0)\Zh_{\omega_j})\,,
\]
for real constants $\alpha_j$.
Using this in (\ref{Z_nu_2_eqtn}) and assessing solvability gives that
\[
-2{\bf E}(\sigma'){\bm \alpha} = {\bf 0}\,, \quad {\bm \alpha} = (\alpha_1,\ldots,\alpha_N)^T\,,
\]
and thus we must have $\sigma' = c$ and ${\bm \alpha} \propto \be$. This is not unexpected as the linear dispersion relation must admit the linear long-wave speed as $\nu \rightarrow 0$, which are equivalent to the characteristics of the WMEs. Without loss of generality make the above proportionality an equivalence for simplicity and thus from (\ref{JC-elements}),
\begin{equation}\label{Z-nu-nu}
Z(\bth,0)_{\nu \nu} = -2 {\bf v}_3-i \sigma''(0)\sum_{j = 1}^N\zeta_j \Zh_{\omega_j}\,,
\end{equation}
Using the results obtained thus far in (\ref{Z_nu_3_eqtn}) and appealing to solvability, it can be seen that the first term vanishes due to the even multiplicity of the zero eigenvalue of ${\bf L}$ and the final term also vanishes. This leaves only
\[
 -3\sigma''(0)\lth\Zh_{\theta_m},\zeta_j\bigg[ {\bf M}(\Zh_{k_j}-c\Zh_{\omega_j})+({\bf J}-c{\bf M})\Zh_{\omega_j}\bigg] \rth = 0\,, m = 1\ldots N\,.
\]
The inner product in fact leads to lead to the system
\[
3\sigma''(0)(\D_\bk{\bf A}+\D_\bw{\bf B}-2c\D_\bw{\bf A})\be = -3 \sigma''(0){\bf E}(c) \be = {\bf 0}\,.
\]
As the characteristic $c$ is assumed simple, this is only true if $\sigma''(0) = 0\,,$. This is expected as the WMEs for the wave $\Zh$ are hyperbolic, and as such it is (modulationally) stable and the dispersion relation $\sigma$ should therefore be real. Overall, we therefore have
\[
Z(\bth,0)_{\nu \nu \nu} = -6i {\bf v}_4-i \sigma'''(0)\sum_{j = 1}^N\zeta_j \Zh_{\omega_j}
\]
Now, in order for the right hand side of (\ref{Z_nu_4_eqtn}) to lie in the range of ${\bf L}$ we require from all previous results that
\[
 6\lth \Zh_{\theta_m},\big({\bf J}-c{\bf M}\big){\bf v}_4\rth+\sigma'''(0)\lth\Zh_{\theta_m},\zeta_j\bigg[ {\bf M}(\Zh_{k_j}-c\Zh_{\omega_j})+({\bf J}-c{\bf M})\Zh_{\omega_j}\bigg] \rth = 0\,,
\]
which is equivalent to the vector system
\[
{\bf T} = -\frac{1}{6}\sigma'''(0)(\D_\bk{\bf A}+\D_\bw{\bf B}-2c\D_\bw{\bf A})\be  = \frac{1}{6}\sigma'''(0){\bf E}'(c)\be\,.
\]
A projection through a left multiplication by $-\be$ gives the scalar quantity appearing as the coefficient of the dispersion term in theorem \ref{ther-paper}:
\[
-\be^T{\bf T} =  -\frac{1}{6}\sigma'''(0)\be^T {\bf E}'(c)\be\,. 
\]
This is to say that the dispersion relation of the long wave model is consistent with the long wave expansion of the original system's dispersion relation (recalling that the $\be^T{\bf E}'(c)\be$ is the coefficient of the time derivative term in mKdV equation outlined in Theorem \ref{ther-paper}), as one expects.

\subsection{Genuine Nonlinearity and its Role in Phase Modulation}\label{sec:GN}
One of the key results of this paper is to connect the notion of linear degeneracy in the Whitham modulation equations associated with a wavetrain and the emergence of the mKdV equation. In order to do so, we recount the notion of genuine nonlinearity in hyperbolic wave equations. Given such an equation of the form
\[
{\bf u}_t+{\bf F}({\bf u}){\bf u}_x = {\bf 0}\,, \quad {\bf u}(x,t) \in \mathbb{R}^n\,, \quad {\bf F} \in \mathbb{R}^n \times \mathbb{R}^n\,,
\]
where ${\bf u}$ represents a state vector,
then we may identify its characteristics $c({\bf u})$ by the standard relation
\[
{\bf F}R_c = c R_c\,,
\]
for the set of right eigenvectors $R_c({\bf u})$. From these notions, we may state the definition of genuine nonlinearity as follows:
\begin{defn}
We then say the evolution associated with the characteristic speed $c$ genuinely nonlinear, as defined by Lax~{\cite{l73}}, if for this speed we have that
\begin{equation}\label{gen-non-defn}
\D_{\bf u} c \cdot R_c \neq 0 \ \forall \ {\bf u}\,,
\end{equation}
where $\cdot$ denotes the standard inner product on vectors.
If a characteristic fails this criterion and instead
\[
\D_{\bf u} c \cdot R_c = 0 \ \forall \ {\bf u}\,,
\]
then it is said to be linearly degenerate.
\end{defn}
In the linearly degenerate regime, neither rarefactions or shocks form and instead contact discontinuities are operational. However it will be necessary to subsequent discussion for a local definition of these properties, as the phase dynamics considers expansions of the systems about a given basic state. To this end, we define the notion of local genuine nonlinearity as follows:
\begin{defn}
Suppose one considers the evolution associated with the characteristic speed $c$ close to some state point ${\bf u}_0$, then we say that the system is locally genuinely nonlinear whenever
\[
\D_{\bf u} c({\bf u_0}) \cdot R_c({\bf u_0}) \neq 0\,.
\]
Whenever the converse is true, we say that the evolution is locally linearly degenerate.
\end{defn}
With the above notions, we will show that the modified KdV equation is a result of the phase dynamics having a local linear degeneracy along one of its characteristic fields.

In the context of the WMEs (\ref{WME-multi}), we have that the characteristics satisfy (\ref{char-defn}) and the system gives the set of right eigenvectors
\[
R_c = \begin{pmatrix}
\be\\
-c\be
\end{pmatrix}\,, \quad {\rm where} \quad {\bf E}(c)\be = {\bf 0}\,.
\]
In order to construct the expression to assess genuine nonlinearity, we use the Jacobi formula for the differentiation of a determinant to find the $k_i$ derivative of $c$ from (\ref{char-defn}):
\[
{\rm tr}\big({\rm adj}({\bf E}) {\bf E}'\big)c_{k_i}+{\rm tr}\bigg({\rm adj}({\bf E})\frac{\partial {\bf E}}{\partial k_i}\bigg) = 0\,, \quad \implies \quad c_{k_i} = -\frac{{\rm tr}\bigg({\rm adj}({\bf E})\frac{\partial {\bf E}}{\partial k_i}\bigg) }{{\rm tr}\big({\rm adj}({\bf E}){\bf E}'\big)}\,.
\]
Similarly, we also have that
\[
c_{\omega_i} = -\frac{{\rm tr}\bigg({\rm adj}({\bf E})\frac{\partial {\bf E}}{\partial \omega_i}\bigg) }{{\rm tr}\big({\rm adj}({\bf E}){\bf E}'\big)}\,.
\]
Then, the relevant condition to determine genuine nonlinearity
\[
\begin{pmatrix}
\D_\bk c\\
\D_\bw c
\end{pmatrix}
\cdot\begin{pmatrix}
\be\\
-c\be
\end{pmatrix} = -\frac{{\rm tr}\big({\rm adj}({\bf E})(\D_\bk{\bf E}-c\D_\bw {\bf E})\be\big) }{{\rm tr}\big({\rm adj}({\bf E}){\bf E}'\big)}
\]
It should be emphasised that in the above, the derivatives $\D_\bk,\,\D_\bw$ on the right hand side do not operate on the $c$ terms in ${\bf E}$. To simplify this, we note that as ${\bf E}$ is singular with simple zero eigenvalue (as $c$ is assumed to be simple) and symmetric we may write its adjugate as
\[
{\rm adj}({\bf E}) = \frac{\mu({\bf E}(c))\be \be^T}{||\be||^2}\,,
\]
where $\mu({\bf E}(c))$ is the product of the remaining nonzero eigenvalues of ${\bf E}(c)$. 
From this, we may then show that
\begin{equation}\label{LinDeg-Multi}
\begin{pmatrix}
\D_\bk c\\
\D_\bw c
\end{pmatrix}
\cdot\begin{pmatrix}
\be\\
-c\be
\end{pmatrix} = -\frac{\mu({\bf E}(c))}{||\be||^2{\rm tr}\big({\rm adj}({\bf E}){\bf E}'\big)}\ \bigg[\be^T(\D_\bk{\bf E}-c\D_\bw {\bf E})(\be,\be) \bigg]\,.
\end{equation}
The term in the square bracket is exactly the coefficient of the quadratic nonlinearity obtained via phase modulation in both the KdV~\cite{r19} and the Two-Way Boussinesq ~\cite{br17}) in the multiphase modulation theory (noting the different sign conventions of the speed $c$), since we can note that
\[
(\D_\bk-c\D_\bw){\bf E} = \D_\bk^2{\bf B}-c(\D_\bk^2{\bf A}+2\D_\bw\D_\bk{\bf A})+c^2(2\D_\bk\D_\bw{\bf A}+\D_\bw^2{\bf B})-c^3\D_\bw^2{\bf A}\,,
\]
and so the multiphase Whitham modulation equations lose genuine nonlinearity precisely when
\begin{equation}\label{lindeg-cond}
\be^T(\D_\bk^2{\bf B}-c(\D_\bk^2{\bf A}+2\D_\bw\D_\bk{\bf A})+c^2(2\D_\bk\D_\bw{\bf A}+\D_\bw^2{\bf B})-c^3\D_\bw^2{\bf A})(\be,\be) = 0\,.
\end{equation}
Thus, a loss of genuine nonlinearity via a local linear degeneracy implies the loss of the quadratic term in the KdV equation and would suggest the emergence of the mKdV, since this contains the cubic nonlinearity one expects to emerge. All that remains is to demonstrate that the phase modulation will lead to the mKdV in the linearly degenerate case, and this will be undertaken in the following section.

\section{Phase Dynamical Reduction to the Modified KdV Equation}
With the abstract framework necessary to undertake the modulation approach outlined, all that remains is to undertake the calculation to demonstrate its emergence. This is to say that we are to prove theorem \ref{ther-paper} and derive the weakly nonlinear dispersive model that arises in this scenario.

The methodology to obtain the modified KdV in this way is to utilise the ansatz
\begin{equation}\label{ansatz}
Z = \Zh\big(\bth+{\bm \Phi}(X,T); \bk + \eps \bq (X,T),\omega-\eps c \bq +\eps^3 \bO(X,T)\big)+\eps^2 W(\bth+{\bm \Phi};\bk,\bw,\eps)\,,
\end{equation}
where $X = \eps x,\, T = \eps^3 t$ and $\eps \ll 1$.
The phase modulation function ${\bm \Phi}$ is defined as the summation of three vector-valued functions
\[
{\bm \Phi} = \bph(X,T)+\eps {\bm \psi}(X,T)+\eps^2{\bm \alpha}(X,T)
\]
and the wavenumber and frequency modulation functions $\bq,\,\bO$ are defined as ${\bm \Phi}_X,\,{\bm \Phi}_T$ respectively. The presence of the $\bq$ term in the frequency modulation is necessary to ensure the phase consistency condition $(\bth+{\bm \Phi}(X,T))_{xt} = (\bth+{\bm \Phi}(X,T))_{tx}$ in light of the moving frame. For convenience in the analysis, we also expand the remainder/correction term $W$ in a simple asymptotic series,
\[
W = W_0+\eps W_1+\eps^2 W_2+\ldots\,,
\]
in order to make the role of $W$ in the analysis clearer. 

The approach is to substitute the ansatz (\ref{ansatz}) into the multisymplectic Euler Lagrange equations (\ref{ELE-MSF}), Taylor expand $Z$ about $\eps = 0$ and solve the resulting system of equations order by order. By using this ansatz and the multisymplectic formalism, the connection between the solvability conditions which arise from the analysis and the conservation laws for the system become obvious and leads to the universal form of the equation. 

This section presents a summary of the order-by-order analysis. There is significant of overlap with previous phase dynamical analyses in this area, so for brevity the details of such overlap will be less detailed and for complete details the reader is referred to this other work.  The most relevant of these are some recent work on modulation in the moving frame for multiple phases~\cite{br17} and a derivation of the modified KdV for two phases in the laboratory frame~\cite{r18}, and it is from these that the subsequent work draws most heavily.

The leading order is satisfied as $\Zh$ solves (\ref{ELE-MSF}), and the first order in $\eps$ satisfied by appealing to (\ref{k-w-deriv}) and the definition of $\bq$. The terms at order $\eps^2$, when simplified, give the system
\[
{\bf L}W_0 = \sum_{j=1}^N (q_j)_X{\bf K}(\Zh_{k_j}-c\Zh_{\omega_j})\,.
\]
By applying the solvability condition (\ref{solvability}) to determine whether the right hand side is in the range of ${\bf L}$, one generates the vector system
\[
{\bf E}(c)\bq_X = {\bf 0}\,.
\]
Thus, for the problem to be solvable at this order, $c$ must be a characteristic of the Whitham modulation equations and $\bq = \be U(X,T)$. This speed is real providing the Whitham modulation equations are hyperbolic, as assumed within the theorem \ref{ther-paper}. The solution for $W_0$ then reads
\[
W_0 = U_X {\bf v}_3\,,
\]
where ${\bf v}_3$ is the third element of the Jordan chain outlined in \S \ref{sec:JC-theory}. 

It is at this point the analysis of this paper diverges from the existing works. The Euler-Lagrange equation at order $\eps^3$, once simplified, reads
\[
\begin{split}
{\bf L}\big(W_1-U_{XX}{\bf v}_4) =& \  \sum_{j=1}^N (\psi_j)_{XX}{\bf K}(\Zh_{k_j}-c\Zh_{\omega_j})\\[2mm]
&+ UU_X\sum_{i=1}^2\bigg[{\bf K}({\bf v}_3)_{\theta_i}-\D^3S(\Zh)({\bf v}_3,\Zh_{k_i}-c\Zh_{\omega_i})\\[2mm]
& +\sum_{j=1}^2{\bf K}(\Zh_{k_ik_j}-c\big(\Zh_{\omega_ik_j}+\Zh_{k_i\omega_j}\big)+c^2\Zh_{\omega_i\omega_j})\bigg]
\end{split}
\]
Appealing to solvability of this equation and manipulations almost identical to previous work~\cite{br20} and leads to the vector system
\[
\big[c^3\D_\bw^2{\bf A}-c^2(2\D_\bk\D_\bw{\bf A}+\D_\bw{\bf B})+c(2\D_\bk\D_\bw{\bf B}+\D_\bk^2{\bf A})-\D_\bk^2{\bf B}\big](\be,\be) - {\bf E}(c)\bps_{XX} = {\bf 0}\,.
\]
This vector system may be solved exactly when the linear degeneracy condition (\ref{lindeg-cond}) holds, as assumed in the premise of the theorem \ref{ther-paper}. In such cases, this imposes that
\begin{equation}\label{psi_defn}
\begin{gathered}
\bps_X = \frac{1}{2}{\bm \kappa}U^2\,, \\[3mm]
 {\rm with} \quad {\bf E}{\bm \kappa} =  \big[c^3\D_\bw^2{\bf A}-c^2(2\D_\bk\D_\bw{\bf A}+\D_\bw{\bf B})+c(2\D_\bk\D_\bw{\bf B}+\D_\bk^2{\bf A})-\D_\bk^2{\bf B}\big](\be,\be)\,.
\end{gathered}
\end{equation}
The result of the analysis at this order is that $W_1$ is given by
\[
W_1 = U_{XX}{\bf v}_4+UU_X \Xi\,, 
\]
where
\[
\begin{split}
{\bf L}\Xi = 
&\sum_{i=1}^2\bigg[\kappa_i{\bf K}(\Zh_{k_i}-c\Zh_{\omega_i})+{\bf K}({\bf v}_3)_{\theta_i}-\D^3S(\Zh)({\bf v}_3,\Zh_{k_i}-c\Zh_{\omega_i})\\[2mm]
& +\sum_{j=1}^2{\bf K}(\Zh_{k_ik_j}-c\big(\Zh_{\omega_ik_j}+\Zh_{k_i\omega_j}\big)+c^2\Zh_{\omega_i\omega_j})\bigg]
\end{split}
\]

The final order of the analysis considered is $\eps^4$, at which the mKdV equation emerges. Simplifying the Euler-Lagrange equation at this order and using (\ref{psi_defn}) results in
\begin{equation}\label{fourth-order}
\begin{split}
{\bf L}\widetilde{W}_2 =&  \ U_T\sum_{i=1}^2\zeta_i\big({\bf J}\Zh_{\omega _i}+{\bf M}\Zh_{k_i}-2c{\bf M}\Zh_{\omega_i}\big)+U_{XXX}{\bf K}{\bf v}_4\\[3mm]
&+U^2U_X\sum_{j=1}^N\bigg\lbrace \frac{1}{2}\kappa_j\big({\bf K}({\bf v}_3)_{\theta_j}-\D^3S(\Zh)({\bf v}_3,\Zh_{k_j}-c\Zh_{\omega_j})\big)\\
&\hspace{3cm}+\zeta_j\big({\bf K}\Xi_{\theta_j}-\D^3S(\Zh)(\Xi,\Zh_{k_j}-c\Zh_{\omega_j} \big)\\
&\hspace{2cm}+\sum_{m=1}^N\bigg[\frac{3}{2}\kappa_j \zeta_m{\bf K}(\Zh_{k_jk_m}-c(\Zh_{k_j\omega_m}+\Zh_{k_m \omega_j})+c^2\Zh_{\omega_j\omega_m})\\
&\hspace{3cm}-\frac{1}{2}\zeta_j\zeta_m\D^3S(\Zh)({\bf v}_3,\Zh_{k_jk_m}-c(\Zh_{k_j\omega_m}+\Zh_{k_m \omega_j})+c^2\Zh_{\omega_j\omega_m})\\
& \hspace{3cm}-\frac{1}{2}\zeta_j\zeta_m\D^4S(\Zh)({\bf v}_3,\Zh_{k_j}-c\Zh_{\omega_j},\Zh_{k_j}-c\Zh_{\omega_j})\\
&\hspace{2.5cm}-\frac{1}{2}\sum_{n=1}^N\zeta_j\zeta_m\zeta_n{\bf K}\bigg(\Zh_{k_jk_mk_n}-c(\Zh_{k_jk_m\omega_n}+\Zh_{k_jk_n\omega_m}+\Zh_{k_mk_n\omega_j})\\
&\hspace{4cm}+c^2(\Zh_{k_j\omega_m\omega_n}+\Zh_{k_m\omega_j\omega_n}+\Zh_{k_n\omega_j\omega_m})-c^3\Zh_{\omega_j\omega_m\omega_n}\bigg)\bigg]\bigg\rbrace\\[3mm]
&+\sum_{i=1}^N\alpha_i{\bf K}(\Zh_{k_i}-c\Zh_{\omega_j})\,.
\end{split}
\end{equation}
The tilde above $W_2$ denotes the fact that all terms at this order which lie in the range of ${\bf L}$ at this order have been absorbed into it. The exact form of these terms does not matter as the analysis terminates at this order, however any analysis at higher orders of $\eps$ would require these. All that remains is to determine the condition for the remaining terms on the right hand side to also lie in the range of ${\bf L}$ , which results in the mKdV equation sought. Its coefficients are generated by appealing to the solvability of the above, and this is what we now generate.

Firstly, applying the solvability condition (\ref{solvability}) to the term involving $U_T$ gives
\[
\begin{split}
\begin{pmatrix}
\lth \Zh_{\theta_1},\sum_{i=1}^2\zeta_i\big({\bf J}\Zh_{\omega _i}+{\bf M}\Zh_{k_i}-2c{\bf M}\Zh_{\omega_i}\big)\rth\\
\vdots\\
\lth \Zh_{\theta_N},\sum_{i=1}^2\zeta_i\big({\bf J}\Zh_{\omega _i}+{\bf M}\Zh_{k_i}-2c{\bf M}\Zh_{\omega_i}\big)\rth
\end{pmatrix}U_T &= \big( 2c\D_\bw{\bf A}-(\D_\bk{\bf A}+\D_\bw{\bf B}) \big)\be U_T\\
&\equiv {\bf E}'(c)\be U_T\,,
\end{split}
\]
The terms involving $\alpha_i$ simply result in $-{\bf E}(c){\bm \alpha}$, as can be seen by its similarity to the expression arising at second order.
Next, we do the same to the term involving the $U_{XXX}$ term, which by the discussion of \S \ref{sec:JC-theory} gives
\[
\begin{pmatrix}
\lth \Zh_{\theta_1},{\bf K}{\bf v}_4\rth\\
\vdots\\
\lth \Zh_{\theta_N},{\bf K}{\bf v}_4\rth
\end{pmatrix} = -\frac{1}{6}\sigma(0)'''{\bf E}'(c)\be U_{XXX}\,,
\]
where $\sigma$ is the linear dispersion relation about the solution $\Zh$.
 The final term to be considered is the cubic nonlinearity, and a significant level of manipulation along the lines of previous works~\cite{br17,r18}. Due to its complexity, the details of these are presented in appendix \ref{sec:cubic} and we simply state here that the solvability condition applied to these terms generates
\[
\begin{split}
&\begin{pmatrix}
\lth \Zh_{\theta_1},\sum_{j=1}^N\bigg\lbrace \frac{1}{2}\kappa_j\big({\bf K}({\bf v}_3)_{\theta_j}\ldots-c^3\Zh_{\omega_j\omega_m\omega_n}\bigg)\bigg]\bigg\rbrace\\
\vdots\\
\lth \Zh_{\theta_N},\sum_{j=1}^N\bigg\lbrace \frac{1}{2}\kappa_j\big({\bf K}({\bf v}_3)_{\theta_j}\ldots-c^3\Zh_{\omega_j\omega_m\omega_n}\bigg)\bigg]\bigg\rbrace\\
\vdots\\
\end{pmatrix}\\[3mm]
& = -\frac{1}{2}\big[c^4\D_\bw{\bf A}-c^3(3\D_\bw^2\D_\bk{\bf A}+\D_\bw^3{\bf B})+c^2(3\D_k\D_\bw^2{\bf B}+3\D_\bk^2\D_\bw{\bf A})\\
&\hspace{2cm}-c(3\D_\bk^2\D_\bw{\bf B}+\D_\bk^3{\bf A})+\D_\bk^3{\bf B}\big](\be,\be,\be)\\
&+\frac{3}{2}\big[c^3\D_\bw^2{\bf A}-c^2(2\D_\bk\D_\bw{\bf A}+\D_\bw{\bf B})+c(2\D_\bk\D_\bw{\bf B}+\D_\bk^2{\bf A})-\D_\bk^2{\bf B}\big](\be,{\bm \kappa})\\[3mm]
& = -\frac{1}{2}(\D_\bk-c\D_\bw)^2{\bf E}(\be,\be,\be)-\frac{3}{2}(\D_\bk-c\D_\bw){\bf E}(\be,{\bm \kappa})\,.
\end{split}
\]
Combining all of these results, the right hand side of (\ref{fourth-order}) lies in the range of ${\bf L}$ providing that
\begin{multline*}
{\bf E}'(c)\be U_T - \frac{1}{2}\bigg[(\D_\bk-c\D_\bw)^2{\bf E}(\be,\be,\be)+3(\D_\bk-c\D_\bw){\bf E}(\be,{\bm \kappa})\bigg]U^2U_X\\
-\frac{1}{6}\sigma(0)'''{\bf E}'(c)\be U_{XXX}-{\bf E}(c){\bm \alpha}_{XX} = {\bf 0}\,.
\end{multline*}
The role of ${\bm \alpha}$ in the analysis is now clear here - without it, we would have $N$ equations for a single unknown, $U$, and so would result in the imposition $U = 0$ and thus no modulation taking place. To remove it from the analysis, and so to find the scalar equation $U$ must satisfy, we multiply the above by $\be$ on the left to project the system in the direction of the kernel of ${\bf E}$. The result of this is the modified KdV equation
\begin{equation}\label{mKdV-result}
\begin{split}
&\alpha U_T+\beta U^2U_X+\gamma U_{XXX} = 0\,,\\[4mm]
{\rm with} \qquad \alpha &= \be^T{\bf E}'(c)\be\,,\\[2mm]
\beta & = -\frac{1}{2}\bigg[\be^T\big((\D_\bk-c\D_\bw)^2{\bf E}(c)\big)(\be,\be,\be)+3\be^T\big((\D_\bk-c\D_\bw){\bf E}(c)\big)(\be,{\bm \kappa})\bigg]\,,\\[2mm]
\gamma &= -\frac{1}{6}\sigma(0)''' \be^T{\bf E}'(c)\be\,.
\end{split}
\end{equation}
The above equation is valid so long as none of the coefficients are zero. We ensure $\alpha \neq 0$ under the assumption in theorem \ref{ther-paper} that the characteristic is of multiplicity one and thus simple, as $\alpha = 0$ is exactly the condition of coalescing characteristics~\cite{br17}. Thus $\gamma \neq 0$ whenever the long wave expansion of the linear dispersion satisfies $\sigma'''(0) \neq 0$, which is generic for dispersive waves and assumed under the assumptions within the theorem. Currently, there is no generic theory to discern in what cases the coefficient of the cubic term vanishes, but it is assumed not to within the theorem. Therefore, the modified KdV equation emerges as the asymptotically valid reduction of (\ref{ELE-MSF}) whenever the Whitham modulation equations are hyperbolic and they are locally linearly degenerate for the chosen characteristic $c$.

\subsection{Weak Linear Degeneracy and the Gardner Equation}
The derivation of the mKdV (\ref{mKdV-result}) relies on the assumption that there is a local linear degeneracy which for the WMEs is equivalent to (\ref{lindeg-cond}).
If it does not hold at all, the KdV equation is expected, however whenever the quantity (\ref{lindeg-cond}) is small, namely of order $\eps$, a suitable balance involving the cubic nonlinearity can be obtained. There are many settings, primarily in the study of internal wave propagation, where this is the case. This results in a quadratic nonlinearity being retained at the final order of the analysis and (\ref{mKdV-result}) instead becomes the Gardner equation,
\begin{equation}\label{gardner-result}
\begin{gathered}
\alpha U_T+\delta UU_X+\beta U^2U_X+\gamma U_{XXX} = 0\,, \\[3mm]
\delta = \big[c^3\D_\bw^2{\bf A}-c^2(2\D_\bk\D_\bw{\bf A}+\D_\bw{\bf B})+c(2\D_\bk\D_\bw{\bf B}+\D_\bk^2{\bf A})-\D_\bk^2{\bf B}\big](\be,\be) \sim \eps\,,
\end{gathered}
\end{equation}
is instead the modulation equation which arises. The key reason the analysis is able to proceed in this case is because the linear system (\ref{psi_defn}) is still solvable in a weak sense, namely that it is solved to leading order with an error of order $\eps$. The projection of this error generates the above quadratic nonlinearity.

\section{Examples of the Theory's Application}\label{sec:examples}
With the abstract result confirmed, we now demonstrate how it may be applied to problems of interest. Namely, we show how nonlinear dispersive models can be constructed using the above result and thus subverting the need for any further asymptotic analysis. The first is a singled phased example to concisely pin down the steps one can take to reach the modified KdV and how one identifies the local linear degeneracy condition required for it to be valid. This will be done for a higher order Nonlinear Schr\"odinger model, which arises in both optical and oceanic settings.  This is then used as a basis to show that the insight gleamed from this example is also applicable to the analysis of Stokes waves~\cite{wlnlw}, which are prevalent across many nonlinear wave systems including water waves~\cite{m82,w67} and viscous fluid conduits~\cite{mh16,jp20}. The second, motivated by recent experiments~\cite{bs14,cdh15,csd19,dsc19}, considers a multilayered shallow water system to demonstrate how a multiple phased relative equilibrium may be treated and how the resulting dispersive reduction can be used to explain the experimental observations.

\subsection{Application to Higher Order NLS}
An illuminating example with a single-phase wavetrain is the higher order Nonlinear Schr\"odinger (NLS) equation. This is given by
\begin{equation}\label{HONLS}
iA_t+\alpha_1A_{xx}-i\alpha_2 A_{xxx}+\beta|A|^2A = 0\,.
\end{equation}
The real constants $\alpha_n$ are  related to the dispersion relation $\omega_0(\kappa)$ of system from which it is derived, namely $\alpha_n = \frac{1}{(n+1)!}\frac{d^{n+1} \omega_0}{d \kappa^{n+1}}$, and $\beta$ relates to the nonlinear correction to it. Conforti et al~\cite{ct13,cbt14,ctmk15,mct14} consider this and related equations to describe the evolution of resonances within optical fibres, but the above NLS equation (sometimes with further nonlinearities) also appears within the study of oceanic Stokes waves~\cite{a00,tkd00}. 

The simplest illustration of the theory of this paper is to investigate the phase dynamics of the genus 0 solution, which is the plane wave
\begin{equation}\label{NLS-sol}
A = A_0e^{i \theta}\,, \quad \mbox{where} \quad |A_0|^2 = \beta^{-1}(\omega+\alpha_1k^2+\alpha_2 k^3)>0\,.
\end{equation}
The conservation law components for this system evaluated on this solution are given by
\begin{equation}\label{cons-laws-NLS}
\mathscr{A}(k,\omega) = \frac{1}{2}|A_0|^2\,, \qquad \mathscr{B}(k,\omega) = k\bigg(\alpha_1+\frac{3 \alpha_2 k}{2}\bigg)|A_0|^2\,,
\end{equation}
which can be used to find the characteristics as
\begin{equation}\label{char-NLS}
c_\pm = 2\alpha_1k+3\alpha_2 k^2\pm \sqrt{-2|A_0|^2\beta(\alpha_1+3 \alpha_2 k)}\,.
\end{equation}
The eigenvector $\be$ associated with each characteristic in this case is simply unity, since there is only a single phase present. Hyperbolicity requires that $\beta(\alpha_1+3\alpha_2 k)<0$, which is a higher order dispersive correction to the typical Benjamin-Feir-Lighthill condition $\alpha_1 \beta<0$, which is recovered for the Stokes wave case $k = 0$. This confirms $\alpha_2$'s secondary effect of stability, however as we will shortly see it has a fundamental role in the characteristics undergoing a local linear degeneracy.

To determine the conditions for the local linear degeneracy, compute the relevant derivatives for the nonlinear term:
\[
\begin{gathered}
\mathscr{A}_{\omega \omega} = \mathscr{A}_{\omega k} = 0\,, \qquad \mathscr{A}_{kk} = \beta^{-1}(\alpha_1+3 \alpha_2 k)\,, \\ 
\mathscr{B}_{kk} = 3 \alpha_2 |A_0|^2+3\beta^{-1} (\alpha_1+3\alpha_2 k)(2\alpha_1k+3 \alpha_2 k^2)\,.
\end{gathered}
\]
Thus, the linear degeneracy condition requires that
\begin{equation}\label{lin-degen-NLS}
-3 c \mathscr{A}_{kk}+\mathscr{B}_{kk} = 3 \alpha_2 \beta |A_0|^2-3(\alpha_1+3\alpha_2 k)\big(c-2\alpha_1k-3 \alpha_2 k^2) = 0\,.
\end{equation}
This occurs when
\[
\alpha_2^2 |A_0|^2 + 2\beta^{-1}(\alpha_1+3 \alpha_2 k)^3 = 0\,, 
\]
and requires the speed whose root is the same sign as $\alpha_2 \beta$ to be chosen, meaning the mKdV equation may only arise for one of the speeds. It is now clear, even in the Stokes wave case of $k=0$ that $\alpha_2$ is required to be nonzero for a local linear degeneracy, highlighting the higher order dispersion's role in this transition.
The vector ${\bm \kappa}$ does not need to be considered for this single phase example, where it is fact zero.

Now that the linear degeneracy condition (\ref{lin-degen-NLS}) has been identified, all that remains is to compute the coefficients. The necessary derivatives of the conservation law components to compute the cubic nonlinearity  are
\[
\mathscr{A}_{\omega \omega \omega} = \mathscr{A}_{\omega \omega k} = \mathscr{A}_{\omega kk} = 0\,, \quad \mathscr{A}_{kkk} = 3 \alpha_2 \beta\,, \quad \mathscr{B}_{kkk} = 12\alpha_2 \beta(2\alpha_1k+3 \alpha_2 k^2)+6 \beta(\alpha_1+3 \alpha_2 k)^2\,.
\]
Thus, the coefficient of the cubic nonlinearity is
\[
-\frac{1}{2}(-4c\mathscr{A}_{kkk}+\mathscr{B}_{kkk})  = \frac{15\alpha_2^2 |A_0|^2}{2(\alpha_1+3 \alpha_2 k)} 
\]
The time derivative coefficient is simply
\[
2c \mathscr{A}_\omega-\mathscr{A}_k-\mathscr{B}_\omega =   \frac{\alpha_2|A_0|^2 }{(\alpha_1+3 \alpha_2 k)}\,.
\]
Finally, we compute linear dispersion relation about this wave by either using a Madelung transform (see \cite{cds12} and references therein) or a Stuart-DiPrima-like analysis~\cite{sd78}, giving
\[
\sigma(\nu) = (2\alpha_1k+3 \alpha_2 k^2)\nu+\alpha_2 \nu^3 \pm \nu\sqrt{-2\beta|A_0|^2(\alpha_1+3 \alpha_2 k)+(\alpha_1+3 \alpha_2 k)^2\nu^2}
\]
The long wave expansion of this is readily computed, and one can find that, using the condition (\ref{lin-degen-NLS}),
\[
\frac{1}{6}\sigma'''(0) =  \frac{3\alpha_2}{4}\,,
\]
Therefore, the modified KdV equation one obtains can be simplified to
\begin{equation}\label{mKdV-NLS}
U_T+\frac{15 \alpha_2}{2}U^2U_X-\frac{3 \alpha_2}{4}U_{XXX} = 0\,.
\end{equation}

There is much about the dynamics of the original wave that may be inferred by the mKdV (\ref{mKdV-NLS}) which one can readily see is the defocussing mKdV.
This implies that all of its periodic solutions are stable~\cite{bhj16}, and these manifest themselves as undulations to the amplitude of the original wave (\ref{NLS-sol}), suggesting the weak formation of wavepackets.
Solitary wave solutions exist for the defocussing mKdV so long as these have a non-zero background, and these correspond to bright and dark solitary waves forming from (\ref{NLS-sol}). 
A family of front solutions however does exist in the defocussing mKdV, which for this problem take the form
\[
U = \pm a_0\tanh\bigg(\sqrt{\frac{5}{3}}a_0(x-vt)\bigg)\,, \quad  v = \frac{5 \alpha_2 a_0^2}{2}\,,
\]
and correspond to smooth shocks in the amplitude of the original wave. 
A full study of these solution families and their effect on the original wave is outside the remit of this paper, but such inferences already demonstrate the level of insight that the theory of this paper can afford regarding the evolution of waves such as (\ref{NLS-sol}) in locally linear degenerate circumstances.

\subsubsection{Connection to Stokes Wave Analyses}
The higher order NLS equation (\ref{HONLS}) provides an informative example under which the computations may be done exactly, however it also provides insight into the phase dynamics of Stokes waves. We will illustrate how this can be done below, demonstrating how the insight of the analysis of the above example mirrors that of weakly nonlinear waves which retain full dispersive information.

Stokes wave solutions are a weakly nonlinear correction to linear waves, leading to corrections to both the wave's amplitude and frequency~\cite{wlnlw}. In doing so, one induces an effective Lagrangian of the form
\begin{equation}\label{Stokes-Lag}
\mathscr{L} = \frac{1}{2}\Omega(k,\omega)a^2-\frac{1}{4}\Gamma a^4+\mathcal{O}(a^6)\,,
\end{equation}
where $a$ is the wave amplitude, which is assumed to be small, $\Omega(k,\omega)$ denotes the linear dispersion relation of the governing equations and $\Gamma$ represents the nonlinear correction to the linear wave's Lagrangian. To the order of the analysis described here, it can be treated as constant, but for more detailed analyses it will vary with the wavenumber $k$. One should note the effects of mean flow have been neglected in the above Lagrangian, which has been done in order to retain parallels to the previous example, but such effects can be important to the evolution of the Stokes wave. An in-depth discussion of these from the perspective of this paper's approach is reserved for future study.

Variations of the Lagrangian (\ref{Stokes-Lag}) with respect to the wave parameters $a,\,k$ and $\omega$ yield expressions for the amplitude as well as the conservation laws we will need to investigate the phase dynamics. Firstly, variations with respect to $a$ lead to the relation which connects the wave parameters to one another:
\[
\delta_a \mathscr{L} = \Omega(k,\omega)a-\Gamma a^3 = 0\,, \quad \Rightarrow \quad a^2 = \Omega \Gamma^{-1}\,.
\]
This variation also allows one to connect $\Gamma$ to the nonlinear frequency correction for the Stokes wave, $\omega_2$ for right moving near-linear waves~\cite{wlnlw}. To do so Taylor expand the dispersion relation about $\omega = -\omega_0(k)$, where $\omega_0$ is the right-moving root of the linear dispersion relation $\Omega$, to show that
\[
\omega = -\omega_0+\frac{\Gamma}{\Omega_\omega(k,-\omega_0)}a^2\,, 
\]
and thus a comparison to the literature gives $\omega_2 = \Gamma \Omega(k,-\omega_0)^{-1}$. Considering the variations of (\ref{Stokes-Lag}) with respect to $k$ and $\omega$ gives the conservation law components
\[
\mathscr{A} = \frac{1}{2}\Omega_\omega a^2\,, \quad \mathscr{B} = \frac{1}{2}\Omega_k a^2\,.
\]
Notice that this recovers exactly (\ref{cons-laws-NLS}) for the case $\Omega = \omega+\alpha_1 k^2+\alpha_2 k^3$, which is expected given the relation between NLS models and Stokes waves.

With the conservation laws determined, we now seek to obtain the characteristics for the case of right-moving near-linear waves, meaning that throughout we will be evaluating the conservation law derivatives at $\omega = -\omega_0$. This leads to the following useful expressions for derivatives of $\Omega$:
\[
\begin{gathered}
\Omega_k(k,-\omega_0) = -\omega_0'\Omega_\omega(k,-\omega_0)\,, \quad \Omega_{\omega k}(k,-\omega_0 = -\omega_0'\Omega_{\omega \omega}(k,-\omega_0)\,, \\[3mm]
 \Omega_{kk}(k,-\omega_0) = -\omega_0''\Omega_\omega(k,-\omega_0)+(\omega_0')^2\Omega_{\omega \omega}(k,-\omega_0)\,,
 \end{gathered}
\]
with the prime denoting derivatives with respect to $k$. With this in mind, we find that the characteristics satisfy
\[
(\Omega_{\omega \omega} a^2+\Omega_\omega^2 \Gamma^{-1}) \ c^2+2(\Omega_{\omega k} a^2+\Omega_\omega \Omega_k \Gamma^{-1}) \ c+\Omega_{kk} a^2+ \Omega_k^2 \Gamma^{-1} = 0\,.
\]
In the small amplitude limit, one is able to find the characteristics in the form of $c = c_0+c_2 a+ \mathcal{O}(a^2)$, and results in the classical nonlinear splitting of the group velocity
\[
c_\pm = \omega_0' \pm \sqrt{\omega_0''\omega_2}a+\mathcal{O}(a^2)\,.
\]
This is of exactly the same form as (\ref{char-NLS}) with $\omega_0 = \alpha_1 k^2+\alpha_2 k^3$ and $\beta = -\omega_2$. For hyperbolicity we require the classical Benjamin-Feir-Lighthill condition of $\omega_0''\omega_2>0$
We now assess the condition for a local linear degeneracy, which once the necessary derivatives of the conservation laws are computed yields the condition
\begin{equation}\label{lin-degen-stokes}
-\frac{1}{2}(\omega_0''' a^2 \pm 3\omega_0'' \omega_2^{-1}\sqrt{\omega_0'' \omega_2} a)\Omega_\omega = 0\,,
\end{equation}
which is identical to (\ref{lin-degen-NLS}) when the previous mentioned choices for $\Omega,\,\omega_0$ and $\beta$ are made, up to a scaling factor. As such, rearranging the above gives the identical condition for a local linear degeneracy,
\[
\omega_2(\omega_0''')^2 a^2-9(\omega_0'')^3 = 0
\]
Thus, the plane wave analysis of the higher order NLS model mirrors that for Stokes waves, and thus such models serve as a good basis to understand Stokes waves in a formulation where the calculations are exact. It also reinforces that the properties of the linear dispersion relation play a substantial role in the nonlinear phase dynamics of the Stokes waves, as initially identified in the previous example.

With the linear degeneracy condition identified, all that remains is to compute the relevant mKdV equation for the Stokes wave in the case where (\ref{lin-degen-stokes}) holds. We start by computing the coefficient of the time derivative term, giving
\[
2c \mathscr{A}_\omega - \mathscr{A}_k-\mathscr{B}_\omega = \pm \Omega_\omega \omega_2^{-1} \sqrt{\omega_0'' \omega_2}a +\mathcal{O}(a^3) = \frac{\omega_0'''\Omega_\omega}{3 \omega_0''} a^2+\mathcal{O}(a^3)\,.
\]
For the dispersive term, we require the linear dispersion relation about the Stokes wave solution (as opposed to the linear dispersion relation for the original system, $\Omega$), which we obtain by using the higher order NLS equation (\ref{HONLS}) as in the previous example, as is the typical approach~\cite{a00,tkd00}. This means the dispersion relation is identical to the previous example and thus
\[
\frac{1}{6}\sigma'''(0) = \frac{\omega_0'''}{8}\,.
\]
Finally, one must compute the coefficient of the cubic nonlinearity. Once simplified using the condition (\ref{lin-degen-stokes}) gives
\[
\begin{gathered}
-\frac{1}{2}(c^4 \mathscr{A}_{\omega \omega \omega}-4c^3 \mathscr{A}_{\omega \omega k}+6c^2\mathscr{A}_{\omega k k}-4c \mathscr{A}_{kkk}+\mathscr{B}_{kkk})= -\frac{1}{2}\bigg(\omega_0^{(iv)}-\frac{5 (\omega_0''')^2}{9\omega_0''}\bigg) \Omega_\omega a^2 +\mathcal{O}(a^3)\,.
\end{gathered}
\]
This is similar to the coefficient obtained for the higher order NLS in the previous example, however there is now a higher order derivative of $\omega_0$ present. It is likely that this additional term would be recovered if further spatial derivatives were included in (\ref{HONLS}). Combining these results gives that the mKdV operational for Stokes waves when (\ref{lin-degen-stokes}) holds is
\begin{equation}
U_T+\frac{1}{2}\bigg(\frac{5 \omega_0'''}{3}-\frac{3\omega^{(iv)}\omega_0''}{\omega_0'''} \bigg)U^2 U_X - \frac{\omega_0'''}{8}U_{XXX} = 0\,.
\end{equation}

This extension of the previous example shows that NLS-type models can yield much of the relevant information one needs to assess the evolution of Stokes waves, with the key difference resulting from the level of dispersive information the Stokes waves inherently possess. This can be remedied simply by adding further derivative terms to the NLS model used, as mentioned prior. However, unlike the mKdV (\ref{mKdV-NLS}), this additional term within the nonlinear coefficient for the Stokes waves case demonstrates that the classification of the mKdV derived can change from focussing to defocussing. This transition depends on the dispersive nature of the system the waves originate from, so a definitive analysis of this mKdV and its effect on the original Stokes is not investigated here. 

\subsection{Application to Stratified Hydrodynamics}
Another particularly illuminating application of the theory of this paper is to stratified fluids. The fact that a modulation-based approach would be operational in such a system is surprising, but it arises from the fact that a set of affine symmetries is present and so the uniform flow solution forms a relative equilibrium. This allows the theory to proceed as described within the paper. A benefit of investigating this system is to make the connection between the characteristics discussed in this paper and the linear long wave speeds widely discussed within the area - in fact, they are the same - and so highlight that these can be used as diagnostic tool to determine the relevant dispersive dynamics.

\begin{figure}
\centering
\includegraphics[width=0.7\textwidth]{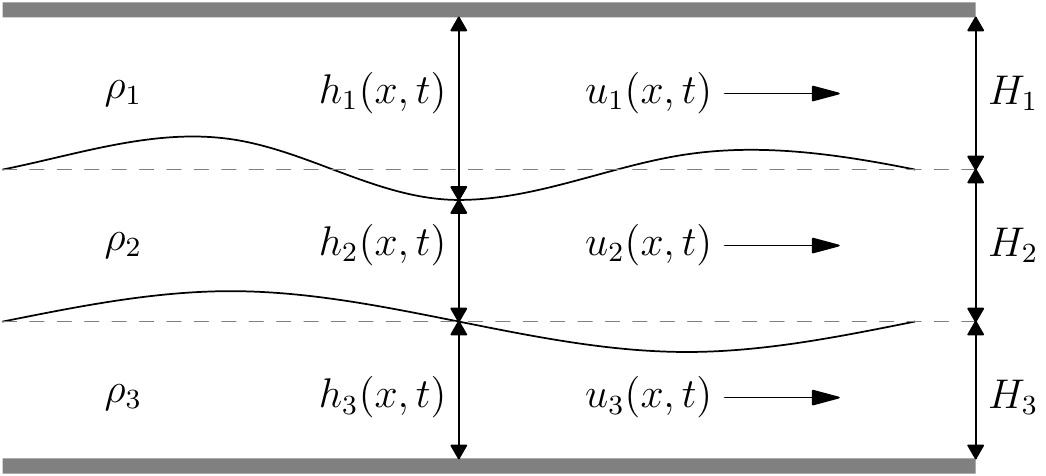}
\caption{A sketch of the three-layered fluid system under consideration with rigid lid.}.
\label{fig:3lf}
\end{figure}

Motivated by the recent experiments on three layered flow~\cite{bs14,cdh15,csd19,dsc19}, we will discuss the shallow water system arising from the Choi-Camassa equations~\cite{bcm20} with linearised dispersion:
\begin{subequations}\label{3-layer-swh}
\begin{align}
&\rho_i\big((h_i)_t+(h_i u_i)_x\big) = 0\,, \quad i = 1,\ldots 3\,,\\[3mm]
 \rho_1\big((u_1)_t+u_1(u_1)_x+g(h_1&+h_2+h_3)_x\big)+P_x = -\frac{\rho_1H_1}{3}(h_1)_{xtt}-\frac{\rho_1H_1}{2}(h_2+h_3)_{xtt}\,,\\[3mm]
  \rho_2\big((u_2)_t+u_2(u_2)_x+g(h_2&+h_3)_x\big)+g \rho_1(h_1)_x+P_x = -\frac{\rho_1H_1}{2}(h_1)_{xtt} \notag\\
  & -\bigg(\frac{\rho_2H_2}{3}+\rho_1H_1\bigg)(h_2)_{xtt}-\bigg(\frac{\rho_2H_2}{2}+\rho_1H_1\bigg)(h_3)_{xtt}\,,\\[3mm]
   \rho_3\big((u_3)_t+u_3(u_3)_x+g(h_3&)_x\big)+g (\rho_1h_1+\rho_2 h_2)_x+P_x = -\frac{\rho_1H_1}{2}(h_1)_{xtt} \notag\\
  &-\bigg(\frac{\rho_2H_2}{2}+\rho_1H_1\bigg)(h_2)_{xtt} -\bigg(\frac{\rho_3H_3}{3}+\rho_2H_2+\rho_1H_1\bigg)(h_3)_{xtt}\,.
\end{align}
\end{subequations}
for layer thicknesses $h_i$, quiescent thicknesses $H_i$ layer density $\rho_i$ and fluid velocity in each layer $u_i$. The layers are labelled from top to bottom, so that the top-most layer is index by 1 and the lowest by 3, meaning that stable stratification requires $\rho_1<\rho_2<\rho_3$. The pressure term $P$ is chosen based on the configuration, namely whether the upper-most surface is free ($P$ is a constant) or whether there is a rigid lid, which is the case for the experimental set-up we wish to consider. This rigid lid imposes the constraint on the thicknesses
\begin{equation}\label{lid-cond}
h_1+h_2+h_3 = H = H_1+H_2+H_3\,.
\end{equation}
One is able to then eliminate the lid pressure by subtracting one momentum equation in (\ref{3-layer-swh}) and one of the thicknesses using (\ref{lid-cond}). Doing so leads to the system of equations
\begin{subequations}\label{3-layer-rigid}
\begin{align}
&\rho_i\big((h_i)_t+(h_i u_i)_x\big) = 0\,, \quad i = 1,3\,,\\[3mm]
&\rho_1\big((H-h_1-h_3)_t+(u_1(H-h_1-h_3)))_x\big) = 0\,.\\[3mm]
  \big(\rho_2u_2-\rho_1 u_1)_t+\bigg(\frac{\rho_2}{2}u_2^2-&\frac{\rho_1}{2}u_1^2+g(\rho_2-\rho_1)(H-h_1)\bigg)_x =& \notag \\
  &\hspace{2.5cm}\frac{\rho_1H_1+\rho_2H_2}{3}(h_1)_{xtt} -\frac{\rho_2H_2}{6}(h_3)_{xtt}\\
   (\rho_3u_3-\rho_2u_2)_t+\bigg(\frac{\rho_3}{2}u_3^2-&\frac{\rho_2}{2}u_2^3 +g(\rho_3-\rho_2) h_3\bigg)_x = \frac{\rho_2H_2}{6}(h_1)_{xtt} -\frac{\rho_2H_2+\rho_3H_3}{3}(h_3)_{xtt}\,.
\end{align}
\end{subequations}
This configuration is sketched in figure \ref{fig:3lf}. The Lagrangian structure emerges once one imposes that the flow is irrotational and introduces the velocity potential in each fluid $\phi_j$ such that $u_j = (\phi_j)_x$. By doing so, this allows one to introduce the Lagrangian
\begin{multline}\label{SWH_Lag}
\mathcal{L} = \iint \bigg[\big(\rho_1(\phi_1)_t-\rho_2(\phi_2)_t\big)h_1+\rho_3(\phi_3)_t-\rho_2(\phi_2)_t\big)h_3\\
+\frac{1}{2}\bigg(\big(\rho_1(\phi_1)_x^2-\rho_2(\phi_2)^2_x\big)h_1+\big(\rho_3(\phi_3)_x^2-\rho_2(\phi_2)^2_x\big)h_3+g(\rho_1-\rho_2)^2h_1^2+g(\rho_3-\rho_2)h_3^2\bigg)\\
-\frac{\rho_1H_1+\rho_2H_2}{6}(h_1)_t^2+\frac{\rho_2H_2}{6}(h_1)_t(h_3)_t-\frac{\rho_2H_2+\rho_3H_3}{6}(h_3)_t^2\\
+H\big(\rho_2(\phi_2)_t+\frac{\rho_2}{2}(\phi_2)_x^2+(\rho_1-\rho_2)h_1\big)\bigg] \ dx dt\,,
\end{multline}
which generates the potential version of (\ref{3-layer-rigid}) as its Euler-Lagrange equations.

The relative equilibrium we study in this example is precisely the uniform flow solution, given by
\[
\phi_i = \theta_i =  U_i x+\omega_i t\,, \quad h_i = H_i\,.
\]
The flow velocities are given by $U_i$ (taking the place of $k_i$ in the theory, to better fit with the literature) and $\omega_i$ represents the Bernoulli head of each flow. Substitution into the Euler-Lagrange equations generated by (\ref{SWH_Lag}) determines the quiescent thicknesses in terms of $U_i,\,\omega_i$.
The vector-valued conservation laws one extracts from this system come from the conservation of mass equations in the shallow water system. Explicitly, we have
\[
{\bf A}(\bU,\bw) = \begin{pmatrix}
\rho_1 H_1(\bU,\bw)\\
\rho_2 H_2(\bU,\bw)\\
\rho_3 H_3(\bU,\bw)
\end{pmatrix}\,,\quad 
{\bf B}(\bU,\bw) = \begin{pmatrix}
\rho_1 H_1 U_1\\
\rho_2 H_2 U_2\\
\rho_3 H_3 U_3
\end{pmatrix}\,.
\]
With the rigid lid constraint (\ref{lid-cond}) and consideration of the static state $U_1 = U_2 = U_3 = 0$, the characteristics which emerge from the determinant condition (\ref{char-defn}) satisfy the biquadratic
\begin{multline}\label{swh-chars}
\big(\rho_1\rho_2H_3+\rho_1\rho_3H_2+\rho_2\rho_3H_1)c^4\\
-\big(\rho_2(\rho_2-\rho_1)H_1H_2+\rho_2(\rho_2-\rho_1)H_1H_3+\rho_1(\rho_3-\rho_2)H_2H_3 \big)c^2\\
+gH_1H_2H_3(\rho_3-\rho_2)(\rho_2-\rho_1) = 0\,,
\end{multline}
which always has real roots~\cite{bcm20}.
This also gives the eigenvector $\be$ as
\[
\begin{split}
\be =& \bigg(\frac{1}{H_1},\frac{\gamma -1}{H_2},-\frac{\gamma}{H_3}\bigg)^T\,,
 \\[4mm] &{\rm with} \quad \gamma = 1+\frac{\rho_1H_2}{\rho_2H_1}-\frac{g(\rho_2-\rho_1)H_2}{\rho_2 c^2} = \bigg( 1+\frac{\rho_3H_2}{\rho_2H_3}-\frac{g(\rho_3-\rho_3)H_2}{\rho_2 c^2} \bigg)^{-1}\,.
 \end{split}
\]
The quantity $\gamma$ represents the ratio between the deflections of the two free surfaces of the problem~\cite[eqtn 2.16]{bcm20}, and it is positive for mode-1 waves (associated with the two largest magnitude roots of (\ref{swh-chars})) and negative for mode-2 (the lowest two magnitude solutions). Therefore, the faster speeds generate waves with the same polarity and slower waves admit waves of opposing polarities.

We are now in a position to assess the criterion for a loss of genuine nonlinearity, which gives
\begin{equation}\label{mKdV-cond-SWH}
\be^T(\D_\bU-c\D_\bw){\bf E}(c)(\be,\be) = \frac{3}{c^2}\bigg(\frac{\rho_1}{H_1^2}+\frac{\rho_2(\gamma-1)^3}{H_2^2}-\frac{\rho_3\gamma^3}{H_3^2}\bigg) = 0\,,
\end{equation}
which is a condition highlighted in Barros et al.~\cite{bcm20} for a loss of the quadratic nonlinearity in the KdV equation arising from this system. It has also appeared in Lamb as a condition arising in the study of conjugate flows~\cite{l00}. Using the density stratification data available from experiments, such as those undertaken by Carr~\cite{cdh15} or Brandt and Shipley~\cite{bs14}, it can be seen that this coefficient is much smaller than others that emerge. This suggests that the Gardner equation (\ref{gardner-result}) is operational for the system being considered, irrespective of how close one is to layer thicknesses where (\ref{mKdV-cond-SWH}) holds.
In either case, whenever the condition (\ref{mKdV-cond-SWH}) is satisfied or small, one is able to define the vector ${\bm \kappa}$ as
\[
{\bm \kappa} = \frac{H_2}{c^2 \rho_2(1-\gamma)}
\begin{pmatrix}
-\frac{3 g (\rho_2-\rho_1)}{\rho_1 H_1^2}+\frac{c^2\rho_2(1-\gamma)}{H_1H_2}\bigg(\frac{\gamma-1}{H_2}-\frac{1}{H_1}\bigg)\\
0\\
-\frac{3 \gamma^3 g (\rho_3-\rho_2)}{\rho_3 H_3^2}+\frac{c^2\rho_2(1-\gamma)}{H_2H_3}\bigg(\frac{\gamma-1}{H_2}+\frac{\gamma}{H_3}\bigg)
\end{pmatrix} \quad ({\rm mod} \ \be )
\]

All that remains is to compute the coefficients for the emergent equation. The most readily available is the coefficient of the time derivative term, giving that
\[
\be^T {\bf E}'(c) \be = -\frac{2}{c}\bigg(\frac{\rho_1}{H_1}+\frac{\rho_2(\gamma-1)^2}{H_2}+\frac{\rho_3\gamma^2 }{H_3} \bigg)\,.
\]
Next, we seek the linear dispersion relation for this system. A simple analysis gives that the linear dispersion relation $\sigma$ satisfies
\begin{small}
\begin{multline}
\bigg[\sigma^2\bigg(\frac{\rho_2}{H_2}+\frac{\rho_1}{H_1}+\frac{\nu^2(\rho_1H_1+\rho_2H_2)}{3}\bigg)-g(\rho_2-\rho_1)\nu^2 \bigg]\bigg[\sigma^2\bigg(\frac{\rho_2}{H_2}+\frac{\rho_3}{H_3}+\frac{\nu^2(\rho_2H_2+\rho_3H_3)}{3}\bigg)-g(\rho_3-\rho_2)\nu^2 \bigg]\\
-\frac{\sigma^4\rho_2^2}{H_2^2}\bigg(1-\frac{\nu^2H_2^2}{6} \bigg)^2 = 0
\end{multline}
\end{small}
By utilising a simple series expansion in $\nu$ for $\sigma$ or via differentiation, one finds that
\[
\frac{1}{6}\sigma(0)''' = - \frac{c\big(\rho_1H_1+\rho_2H_2(1+\gamma+\gamma^2)+\rho_3 H_3 \gamma^2 \big)}{6\big(\frac{\rho_1}{H_1}+\frac{\rho_2}{H_2}(1-\gamma)^2 +\frac{\rho_3 \gamma^2}{H_3}\big)}\,.
\]
The cubic nonlinearity is found in parts. The first part calculated is the term involving the third derivative of ${\bf E}$:
\[
\be^T(\D_\bU-c\D_\bw)^3{\bf E}(\be,\be,\be) = -3 \bigg[\frac{1}{g(\rho_2-\rho_1)}\bigg(\frac{\rho_2(\gamma-1)^2}{H_2^2}-\frac{\rho_1}{H_1^2}\bigg)^2+\frac{1}{g(\rho_3-\rho_2)}\bigg(\frac{\rho_2(\gamma-1)^2}{H_2^2}-\frac{\rho_3 \gamma^2}{H_3^2} \bigg)^2\bigg]\,.
\]
The second term necessary for this computation is
\begin{multline}
3(\D_\bU-c\D_\bw){\bf E}(\bkap,\be) = -3\bigg[\frac{\rho_1H_2}{g\rho_2 H_1(1-\gamma)(\rho_2-\rho_1)}\bigg(\frac{\rho_2(\gamma-1)^2}{H_2^2}+\frac{2 \rho_2(\gamma-1)}{H_1H_2}-\frac{3\rho_1}{H_1^2}\bigg)^2\\-\frac{\gamma \rho_3H_2}{g\rho_2H_2(1-\gamma)(\rho_3-\rho_2)}\bigg(\frac{\rho_2(\gamma-1)^2}{H_2^2}-\frac{2 \rho_2(\gamma-1)\gamma}{H_2H_3}-\frac{3\rho_3 \gamma^2}{H_3^2}\bigg)^2\bigg]
\end{multline}
These may be combined and simplified using (\ref{mKdV-cond-SWH}) to give
\[
\begin{split}
\be^T\bigg((\D_\bU-c\D_\bw)^2{\bf E}(c)\bigg)(\be,\be,\be)&+3\be^T\bigg((\D_\bU-c\D_\bw){\bf E}(c)\bigg)(\be,{\bm \kappa})\\
=&\frac{3}{2c^4}\bigg[\frac{9 H_2}{\rho_2}\bigg(\frac{\rho_3\gamma^2}{H_3^2}-\frac{\rho_2(1-\gamma)^2}{H_2^2} \bigg) \bigg(\frac{\rho_2(1-\gamma)^2}{H_2^2}-\frac{\rho_1}{H_1^2} \bigg)\\
&\qquad+4 \bigg(\frac{\rho_1}{H_1^3}+\frac{\rho_2(1-\gamma)^4}{H_2^3}+\frac{\rho_3\gamma^4}{H_3^3} \bigg)\bigg]
\end{split}
\]
Thus, the Gardner equation which emerges for the three-layered system (\ref{3-layer-rigid}) is
\begin{align}\label{mkdv-swh}
\begin{split}
& \hspace{3cm}U_T+\alpha UU_X+\beta U^2U_X+\gamma U_{XXX} = 0\,,\\[5mm]
{\rm with} \quad \alpha &=\frac{3\bigg(\frac{\rho_3\gamma^3}{H_3^2}+\frac{\rho_2(1-\gamma)^3}{H_2^2}-\frac{\rho_1}{H_1^2}\bigg)}{2 c \bigg(\frac{\rho_1}{H_1}+\frac{\rho_2(1-\gamma)^2}{H_2}+\frac{\rho_3 \gamma^2}{H_3}\bigg)}\\[3mm]
\beta  &=  -\frac{3\bigg[\frac{9 H_2}{\rho_2}\bigg(\frac{\rho_3\gamma^2}{H_3^2}-\frac{\rho_2(1-\gamma)^2}{H_2^2} \bigg) \bigg(\frac{\rho_2(1-\gamma)^2}{H_2^2}-\frac{\rho_1}{H_1^2} \bigg)+4 \bigg(\frac{\rho_1}{H_1^3}+\frac{\rho_2(1-\gamma)^4}{H_2^3}+\frac{\rho_3\gamma^4}{H_3^3} \bigg)\bigg]}{8c^3\bigg(\frac{\rho_1}{H_1}+\frac{\rho_2(1-\gamma)^2}{H_2}+\frac{\rho_3 \gamma^2}{H_3}\bigg)}\,,\\[3mm]
\beta &= \frac{c\big(\rho_3 H_3 \gamma^2 +\rho_2 H_2(\gamma^2+\gamma+1)+\rho_1 H_1\big)}{6\bigg(\frac{\rho_1}{H_1}+\frac{\rho_2(1-\gamma)^2}{H_2}+\frac{\rho_3 \gamma^2}{H_3}\bigg)}\,.
\end{split}
\end{align}

\begin{figure}[t!]
\centering
\begin{subfigure}{0.45 \textwidth}
\includegraphics[width = \textwidth]{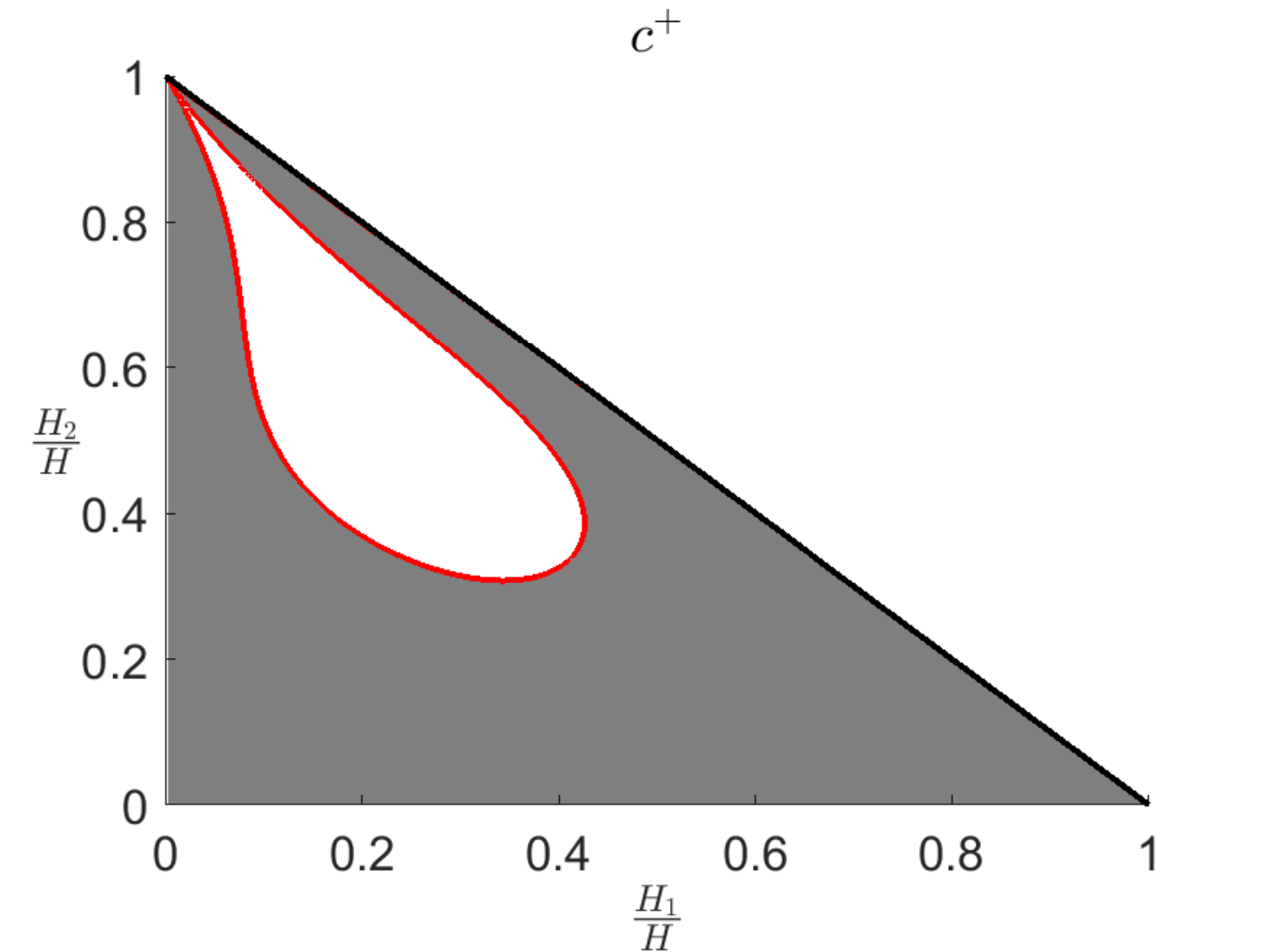}
\end{subfigure}
\begin{subfigure}{0.45 \textwidth}
\includegraphics[width = \textwidth]{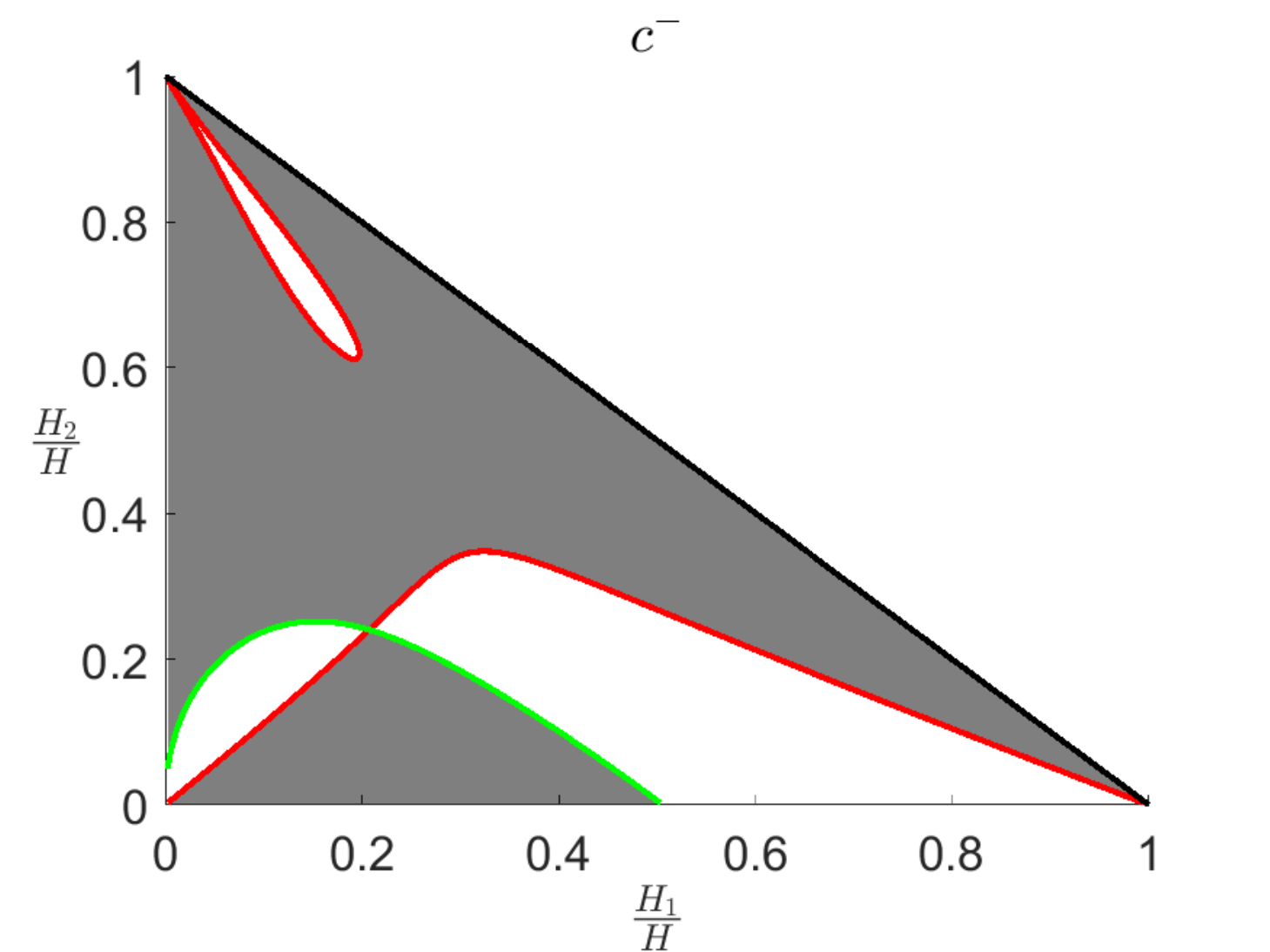}
\end{subfigure}\\
\begin{subfigure}{0.8 \textwidth}
\includegraphics[width = \textwidth]{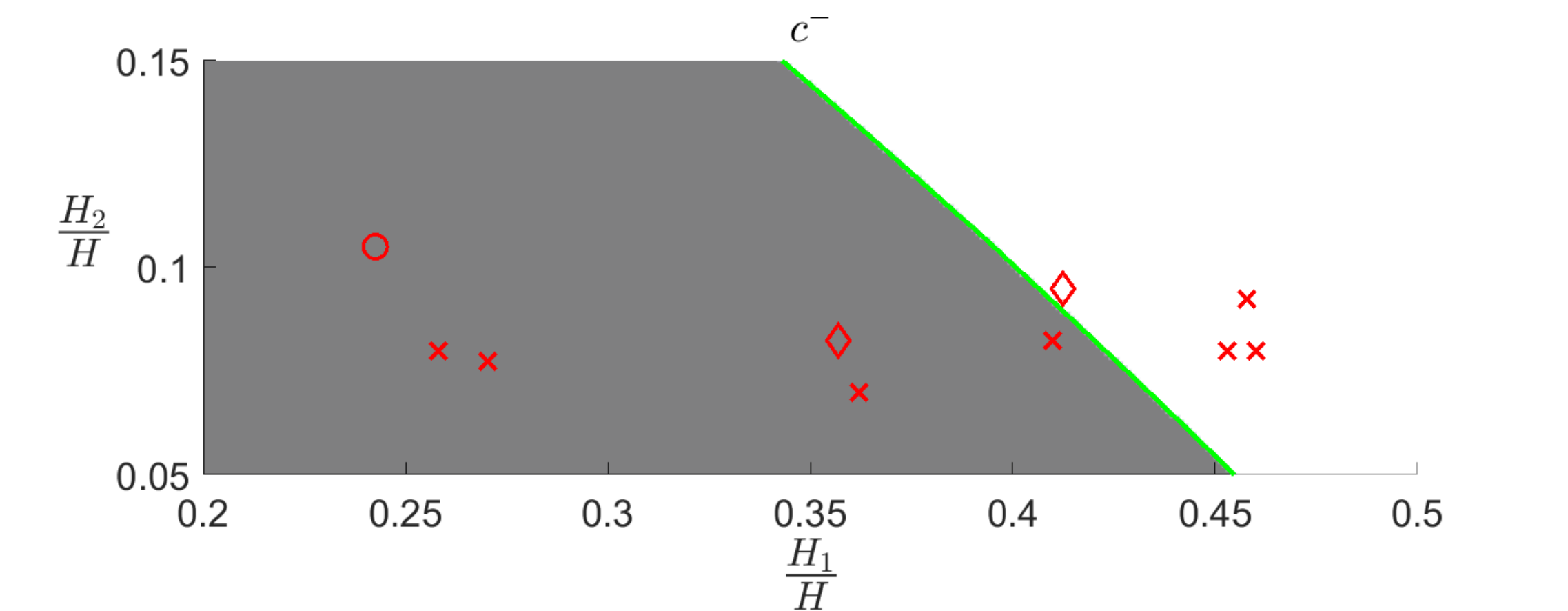}
\end{subfigure}
\caption{A graphical representation of how the polarity of the Gardner equation changes as the thicknesses vary (a) the mode-1 and (b) mode-2 waves for the stratification data in Carr et al. . The red and green lines denote the boundaries across which the cubic and dispersive coefficients change signs respectively. White areas denote focussing regime whereas grey areas signify the defocussing case. Figure (c) includes data points corresponding to low-amplitude wave observations from Carr's experiments, with crosses denoting stable, diamonds marginally stable and circles denoting unstable waves.}
\label{fig:contours}
\end{figure}

The three-layered system (\ref{3-layer-rigid}) has a large parameter space on which the coefficients of (\ref{mkdv-swh}) greatly depend, so fully exploring it in detail is difficult. However, we can use parameters informed by experiments such as Brandt and Shipley~\cite{bs14} and Carr et al.~\cite{cdh15} to explore the nature of the dynamics of (\ref{mkdv-swh}) and assess how the predictions of this equation agree with what is observed in these experiments. Figure \ref{fig:contours} illustrates how this can be achieved for data from the latter work. We see that for the symmetric configurations, where $H_1 = H_3$ and $H_2$ is relatively small, the Gardner equation (\ref{mkdv-swh}) is focussing and all solitary waves are found to be stable~\cite{kn10}. This agrees with what has been experimentally shown for small amplitude waves, for which the Gardner equation is limited to. As this symmetry is broken and the middle layer is offset, as explored in Carr et al.~\cite{cdh15}, the Gardner equation changes from focussing to defocussing type and the stability of solitary waves instead depends on a relationship between the solitary wave amplitude and its speed. This appears to agree with Carr's observations for low amplitude mode-2 waves, where low amplitude waves destabilised as the offset of the middle layer was increased. The quantitative level of agreement for the marginal stability plane between the Gardner equation and the low amplitude waves from experiments, as given by the relation
\[
1-M \frac{c}{a} = 0\,,
\]
for nondimensional wave amplitude $a$, wavespeed $c$ and coefficient $M$ determining the slope. For the Gardner equation $M = 6$ and Carr's data, although there are only two low amplitude data points, suggests $M = 6.79$ for such waves.
Thus it would appear that the Gardner equation gives a good qualitative picture of the dynamics of the three-layered systems that are experimentally considered, and a full quantitative assessment of this descriptive ability is reserved for future study.

This is not to say that the derived Gardner equation gives complete picture of the dynamics, despite its successes and widespread use in internal wave modelling. This insight should only be applicable for sufficiently small amplitude structures in (\ref{3-layer-swh}), where (\ref{mkdv-swh}) is applicable, and therefore cannot explain why internal waves destabilise at larger amplitudes, although it will likely apply to the smaller solitary waves resulting from the fission process observed. It also is unable to explain the observed asymmetry which arises from these experiments. It additionally will not be expected to fully characterise conjugate flow states correctly in all cases~\cite{l00}. For a more comprehensive investigation of these situations, one should instead use strongly nonlinear models to improve accuracy and extend the validity of long wave models.

\section{Concluding Remarks}\label{sec:con}
In this paper, we have connected the notion of genuine nonlinearity to the weakly nonlinear dispersive behaviour of the system, namely that local linear degeneracy signals the emergence of the modified KdV equation. This allows one to use the linear quantity, the characteristic, as a diagnostic for the behaviour of the nonlinear system. Moreover, quantities available from the linear theory form its coefficients and so the mKdV may be constructed simply from a linear analysis of the original wavetrain. The nonlinear equation which results from the analysis can then be used for primarily qualitative insight into the system's evolution at such degeneracy points, however there is some evidence that it may also yield quantitative insight. 

\begin{figure}
\centering
\includegraphics[width=0.6 \textwidth, trim ={5cm 17cm 4cm 4cm}, clip=true]{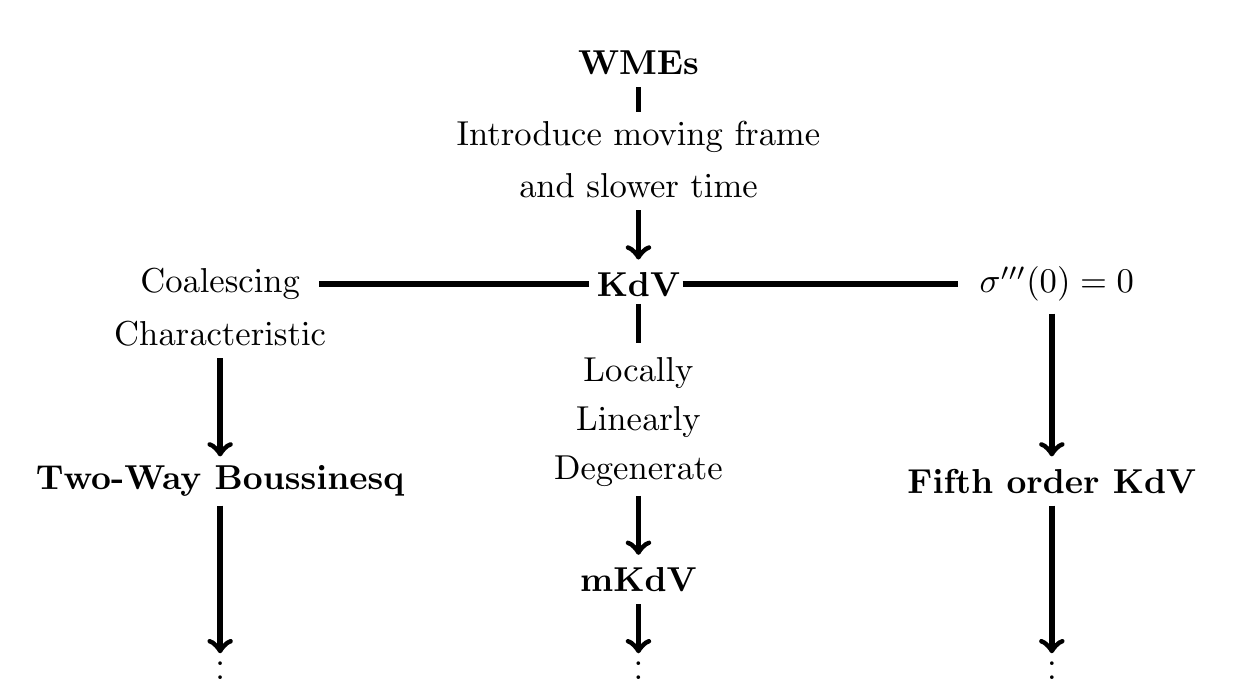}
\caption{A summary of the flow of logic used to determine the appropriate phase dynamical equation for a given nonlinear wave.}
\label{fig:flowchart}
\end{figure}

A consequence of the universality of the resulting phase dynamical equation is that one is able to characterise the dispersive dynamics through conditions imposed on the quantities $c$ and $\sigma$ which one readily obtains via linear analyses. As a result, starting from the WMEs one can systematically identify the most suitable dispersive long-wave model simply by assessing which of the relevant properties the characteristic $c$ and linear dispersion relation $\sigma$ satisfy. Combined with connections made in previous works~\cite{r19,br20} this essentially turns the phase modulation analysis into a flow chart-like process. This is visualised in figure \ref{fig:flowchart}, demonstrating the connections between each of these well-established equation and the conditions required on the respective linear quantity. Further, it suggests that when such conditions are combined, much more complex phase dynamics should be expected. For example, when dispersion is weak and a given characteristic is locally linearly degenerate, the appropriate phase dynamical equation should resemble an extended version of the KdV equation. Moreover, by combining the linear degeneracy of this paper with a double characteristic signals the emergence of a modified version of the two-way Boussinesq equation~\cite{rat-thesis}.

There are several avenues for future work based on these results. One of the most natural directions to take is to use this approach to investigate other nonlinear wave systems to discern the insight the theory offer, such as a more in-depth analysis of the Stokes wave example for particular systems including the water wave problem~\cite{w67}. Most importantly, the interaction between waves and their mean flow have a highly nonlinear interplay in such systems and so the ideas of this paper and preceding work will likely shed significant insight into this coupling. 

Another prospective direction concerns itself with the ``infinite'' phase limit of Whitham modulation theory. The discussion within this work has dealt with finitely many phases, so that the formulation of conservation laws, characteristics and solvability conditions involves only linear algebraic constructions. However, there are examples where the family of relative equilibria depends continuously on a variable, which in essence makes it 'infinite' phased. Such cases arise in the study of continuously stratified water waves~\cite{g77,gos97,gpt97} and wave fields involving a whole spectrum of wavenumbers~\cite{owe12,w75}. In such cases, we expect there to be a spectrum of characteristics $c$ in play, and it therefore isn't clear how the notion of local linear degeneracy will generalise in these contexts but should still lead to the mKdV emerging as it has been shown to for internal waves.

\section*{Acknowledgements}
The author would like to acknowledge the many helpful discussions with Gennady El, Tom Bridges and Ricardo Barros whilst undertaking the work within this paper. The author also deeply thanks Melanie Watson for her support and encouragement throughout the writing of this manuscript.

\appendix

\section{Calculation of the Cubic Coefficient}\label{sec:cubic}
Due to the cumbersome nature of the cubic coefficient's calculation, we undertake it here in an appendix. In short, we wish to try and write the inner product
\begin{equation}\label{Cubic-term-raw-app}
\begin{split}
&\left< \hspace{-2mm} \left<  \Zh_{\theta_i},\sum_{j=1}^N\bigg[ \frac{1}{2}\kappa_j\big({\bf K}({\bf v}_3)_{\theta_j}-\D^3S(\Zh)({\bf v}_3,\Zh_{k_j}-c\Zh_{\omega_j})\big) \right. \right. \\
&\hspace{3cm}+\zeta_j\big({\bf K}\Xi_{\theta_j}-\D^3S(\Zh)(\Xi,\Zh_{k_j}-c\Zh_{\omega_j}) \big)\\
&\hspace{2cm}+\sum_{m=1}^N\big(\frac{3}{2}\kappa_j \zeta_m{\bf K}(\Zh_{k_jk_m}-c(\Zh_{k_j\omega_m}+\Zh_{k_m \omega_j})+c^2\Zh_{\omega_j\omega_m})\\
&\hspace{3cm}-\frac{1}{2}\zeta_j\zeta_m\D^3S(\Zh)({\bf v}_3,\Zh_{k_jk_m}-c(\Zh_{k_j\omega_m}+\Zh_{k_m \omega_j})+c^2\Zh_{\omega_j\omega_m})\\
& \hspace{3cm}-\frac{1}{2}\zeta_j \zeta_m\D^4S(\Zh)({\bf v}_3,\Zh_{k_j}-c\Zh_{\omega_j},\Zh_{k_j}-c\Zh_{\omega_j})\\
&\hspace{2cm}+\frac{1}{2}\sum_{n=1}^N{\bf K}(\Zh_{k_jk_mk_n}-c\Zh_{k_jk_m\omega_n}+\Zh_{k_jk_n\omega_m}+\Zh_{k_mk_n\omega_j})\\
&\hspace{3cm}+\left. \left. c^2(\Zh_{k_j\omega_m\omega_n}+\Zh_{k_m\omega_j\omega_n}+\Zh_{k_n\omega_j\omega_m})-c^3\Zh_{\omega_j\omega_m\omega_n}\bigg)\bigg]\right> \hspace{-2mm}\right>
\end{split}
\end{equation}
in terms of derivatives of the conservation laws. We will ultimately show that this inner product leads to the vector term
\begin{equation}\label{end-result-cubic}
-\frac{1}{2}\bigg((\D_\bk - c\D_\bw)^4{\bf E}(c)(\be,\be,\be)+3(\D_\bk-c\D_\bw)^3{\bf E}(c)(\be,{\bm \kappa})\bigg)\,.
\end{equation}
We do this in stages, as in \cite{rat-thesis,r18}, and will require use to use further derivatives of the basic state. For example, for some of the manipulation we will use
\[
{\bf L}(\Zh_{\theta_i k_j}-c\Zh_{\theta_i\omega_j}) = {\bf K}\Zh_{\theta_i \theta_j}-\D^3S(\Zh)(\Zh_{\theta_i},\Zh_{k_j}-c\Zh_{\omega_j})\,,
\]
to simplify the inner product, and this relation can be obtained simply by differentiation of (\ref{MJS-wavetrain}) with respect to $\theta_i$, then either $k_j,\omega_j$ and combining the results. Further relations of this nature can be obtained in a similar fashion but are not documented here.

We manipulate, starting with the terms involving $\Xi$:
\[
\begin{split}
&\sum_{j=1}^N\zeta_j\lth\Zh_{\theta_i},{\bf K}\Xi_{\theta_j}-\D^3S(\Zh)(\Xi,\Zh_{k_j}-c\Zh_{\omega_j}) \rth = \sum_{j=1}^N\zeta_j\lth{\bf L}(\Zh_{\theta_i k_j}-c\Zh_{\theta_i \omega_j}),\Xi \rth\\[2mm]
& = \sum_{j=1}^N\zeta_j\left< \hspace{-2mm} \left<\Zh_{\theta_i k_j}-c\Zh_{\theta_i \omega_j}, \sum_{m=1}^2\bigg[\kappa_m{\bf K}(\Zh_{k_m}-c\Zh_{\omega_m})\right. \right.\\
& \hspace{2cm}+\zeta_m \bigg({\bf K}({\bf v}_3)_{\theta_m}-\D^3S(\Zh)({\bf v}_3,\Zh_{k_m}-c\Zh_{\omega_m}) \\
& \hspace{2cm}\left. \left. +\sum_{n=1}^2\zeta_n{\bf K}(\Zh_{k_mk_n}-c\big(\Zh_{\omega_mk_n}+\Zh_{k_m\omega_n}\big)+c^2\Zh_{\omega_m\omega_n})\bigg)\bigg] \right> \hspace{-2mm} \right>\\[2mm]
& = \red{\sum_{j,m,n=1}^N \zeta_j\zeta_m \zeta_n\lth \Zh_{\theta_i k_j}-c\Zh_{\theta_i \omega_j},{\bf K}(\Zh_{k_mk_n}-c\big(\Zh_{\omega_mk_n}+\Zh_{k_m\omega_n}\big)+c^2\Zh_{\omega_m\omega_n})\rth}\\
& \hspace{1cm}+\blue{\sum_{j,m=1}^N \zeta_j \kappa_m \lth \Zh_{\theta_i k_j}-c\Zh_{\theta_i \omega_j},{\bf K}(\Zh_{k_m}-c\Zh_{\omega_m})\rth}\\
&\hspace{2cm}+\sum_{j,m = 1}^N\zeta_j \zeta_m\lth \Zh_{\theta_i k_j}-c\Zh_{\theta_i \omega_j},({\bf K}({\bf v}_3)_{\theta_m}-\D^3S(\Zh)({\bf v}_3,\Zh_{k_m}-c\Zh_{\omega_m})\rth\,.
\end{split}
\]
We colour code the terms which require no further manipulation, with red terms contributing to the first term in (\ref{end-result-cubic}) and blue the second. In subsequent lines we contract these term by writing them as \textsc{\red{red}},\,\textsc{\blue{blue}} respectively.
We can then combine these terms with those in (\ref{Cubic-term-raw-app}) involving ${\bf v}_3$:
\[
\begin{split}
&\sum_{j=1}^N\zeta_j\lth\Zh_{\theta_i},{\bf K}\Xi_{\theta_j}-\D^3S(\Zh)(\Xi,\Zh_{k_j}-c\Zh_{\omega_j}) \rth\\
&+\frac{1}{2}\sum_{j=1}^N \bigg[\kappa_j\lth\Zh_{\theta_i},{\bf K}({\bf v}_3)_{\theta_j}-\D^3S(\Zh)({\bf v}_3,\Zh_{k_j}-c\Zh_{\omega_j})\rth\\
&-\zeta_j\sum_{m=1}^N\zeta_m\lth\Zh_{\theta_i},\D^4S(\Zh)({\bf v}_3,\Zh_{k_j}-c\Zh_{\omega_j},\Zh_{k_j}-c\Zh_{\omega_j})\\
&+\D^3S(\Zh)({\bf v}_3,\Zh_{k_jk_m}-c(\Zh_{k_j\omega_m}+\Zh_{k_m \omega_j})+c^2\Zh_{\omega_j\omega_m})\rth\bigg]\\[2mm]
& =\mbox{ \textsc{\red{red}}}+\mbox{\textsc{\blue{blue}}}+\frac{1}{2}\sum_{j=1}^N \kappa_j \lth {\bf L}(\Zh_{\theta_i k_j}-c\Zh_{\theta_i \omega_j}),{\bf v}_3\rth\\
&+\sum_{j,m=1}^N\zeta_j\zeta_m \lth {\bf L}(\Zh_{\theta_i k_j k_m}-c(\Zh_{\theta_i k_j \omega_m}+\Zh_{\theta_i k_m \omega_j})+c^2\Zh_{\theta_i \omega_j \omega_m}),{\bf v}_3\rth\\[2mm]
& =\mbox{ \textsc{\red{red}}}+\mbox{\textsc{\blue{blue}}}+\blue{\frac{1}{2}\sum_{j,m=1}^N \kappa_j\zeta_m \lth (\Zh_{\theta_i k_j}-c\Zh_{\theta_i \omega_j}),{\bf K}(\Zh_{k_m}-c\Zh_{\omega_m})\rth}\\
&+\red{\sum_{j,m,n=1}^N\zeta_j\zeta_m\zeta_n \lth (\Zh_{\theta_i k_j k_m}-c(\Zh_{\theta_i k_j \omega_m}+\Zh_{\theta_i k_m \omega_j})+c^2\Zh_{\theta_i \omega_j \omega_m}),{\bf K}(\Zh_{k_m}-c\Zh_{\omega_m})\rth}
\end{split}
\]
The remaining terms in (\ref{Cubic-term-raw-app}) contribute to the red terms. We now gather the terms of each color, and the simplest of the two are the blue terms. Collecting the powers of $c$ together, one can show that
\[
\begin{split}
\sum_{j,m=1}^N &\zeta_j \kappa_m \bigg[\lth \Zh_{\theta_i k_j}-c\Zh_{\theta_i \omega_j},{\bf K}(\Zh_{k_m}-c\Zh_{\omega_m})\rth\\
&\hspace{3cm}+\frac{1}{2}\sum_{j,m=1}^N \kappa_j\zeta_m \lth (\Zh_{\theta_i k_j}-c\Zh_{\theta_i \omega_j}),{\bf K}(\Zh_{k_m}-c\Zh_{\omega_m})\rth\bigg]\\[2mm]
&= \frac{3}{2}\sum_{j,m=1}^N \zeta_j \kappa_m \bigg[-\frac{1}{2}c^3\partial_{\omega_j \omega_m}\lth \Zh_{\theta_i},{\bf M}\Zh\rth+c^2\bigg(\partial_{k_j \omega_m}\lth \Zh_{\theta_i},{\bf M}\Zh\rth+\frac{1}{2}\partial_{\omega_j \omega_m}\lth \Zh_{\theta_i},{\bf J}\Zh\rth \bigg)\\
& \hspace{2cm} -c\bigg(\frac{1}{2}\partial_{k_j} \partial_{k_m}\lth \Zh_{\theta_i},{\bf M}\Zh\rth+\partial_{k_j}\partial_{ \omega_m}\lth \Zh_{\theta_i},{\bf J}\Zh\rth \bigg)+\frac{1}{2}\partial_{k_j}\partial_{k_m}\lth \Zh_{\theta_i},{\bf J}\Zh\rth\bigg]\\[2mm]
& = -\frac{3}{2} \sum_{j,m=1}^N \zeta_j \kappa_m \bigg(-c^3 \partial_{\omega_j \omega_m}\mathscr{A}_i+c^2(2\partial_{k_j}\partial_{\omega_m}\mathscr{A}_i+\partial_{\omega_j}\partial_{\omega_m}\mathscr{B}_i)\\
& \hspace{2cm}-c(\partial_{k_j}\partial_{k_m}\mathscr{A}_i+\partial_{k_j}\partial_{\omega_m}\mathscr{B}_i)+\partial_{k_j}\partial_{k_m}\mathscr{B}_i\bigg)\,.
\end{split}
\]
This is exactly the index form of the second term in (\ref{end-result-cubic}).
Then by gathering these red terms, we can also show that
\[
\begin{split}
\sum_{j,m,n=1}^N& \zeta_j\zeta_m \zeta_n \bigg[\lth \Zh_{\theta_i k_j}-c\Zh_{\theta_i \omega_j},{\bf K}(\Zh_{k_mk_n}-c\big(\Zh_{\omega_mk_n}+\Zh_{k_m\omega_n}\big)+c^2\Zh_{\omega_m\omega_n})\rth\\
&+\bigg\langle \hspace{-2mm} \bigg\langle (\Zh_{\theta_i k_j k_m}-c(\Zh_{\theta_i k_j \omega_m}+\Zh_{\theta_i k_m \omega_j})+c^2\Zh_{\theta_i \omega_j \omega_m}),{\bf K}(\Zh_{k_m}-c\Zh_{\omega_m})\rth\\
&+\frac{1}{2}\lth \Zh_{\theta_i},{\bf K}(\Zh_{k_jk_mk_n}-c\Zh_{k_jk_m\omega_n}+\Zh_{k_jk_n\omega_m}+\Zh_{k_mk_n\omega_j})\\
&\hspace{3cm}+ c^2(\Zh_{k_j\omega_m\omega_n}+\Zh_{k_m\omega_j\omega_n}+\Zh_{k_n\omega_j\omega_m})-c^3\Zh_{\omega_j\omega_m\omega_n}\bigg)\bigg\rangle \hspace{-2mm} \bigg\rangle  \bigg]\\[3mm]
&= \frac{1}{4}\sum_{j,m,n=1}^N c^4 \bigg(\partial_{\omega_j}\partial_{\omega_m}\partial_{\omega_n}\lth\Zh_{\theta_i},{\bf M}\Zh\rth\\
& -c^3\bigg(3\partial_{k_j}\partial_{\omega_m}\partial_{\omega_n}\lth\Zh_{\theta_i},{\bf M}\Zh\rth +\partial_{\omega_j}\partial_{\omega_m}\partial_{\omega_n}\lth\Zh_{\theta_i},{\bf J}\Zh\rth \bigg)\\
&+c^2 \bigg(3\partial_{k_j}\partial_{k_m}\partial_{\omega_n}\lth\Zh_{\theta_i},{\bf M}\Zh\rth +3\partial_{k_j}\partial_{\omega_m}\partial_{\omega_n}\lth\Zh_{\theta_i},{\bf J}\Zh\rth \bigg)
\\
&-c\bigg(\partial_{k_j}\partial_{k_m}\partial_{k_n}\lth\Zh_{\theta_i},{\bf M}\Zh\rth +3\partial_{k_j}\partial_{k_m}\partial_{\omega_n}\lth\Zh_{\theta_i},{\bf J}\Zh\rth \bigg)+\partial_{k_j}\partial_{k_m}\partial_{k_n}\lth\Zh_{\theta_i},{\bf J}\Zh\rth \\[3mm]
& = -\frac{1}{2}\sum_{j,m,n=1}^N \zeta_j\zeta_m \zeta_n \bigg(c^4 \partial_{\omega_j}\partial_{\omega_m}\partial_{\omega_n}\mathscr{A}_i-c^3(3\partial_{k_j}\partial_{\omega_m}\partial_{\omega_n}\mathscr{A}_i+\partial_{\omega_j}\partial_{\omega_m}\partial_{\omega_n}\mathscr{B}_i)\\
&+c^2(3\partial_{k_j}\partial_{k_m}\partial_{\omega_n}\mathscr{A}_i+3\partial_{k_j}\partial_{\omega_m}\partial_{\omega_n}\mathscr{B}_i)-c(\partial_{k_j}\partial_{k_m}\partial_{k_n}\mathscr{A}_i+3\partial_{k_j}\partial_{k_m}\partial_{\omega_n}\mathscr{B}_i)+\partial_{k_j}\partial_{k_m}\partial_{k_n}\mathscr{B}_i\bigg)\,.
\end{split}
\]
This is the index form of the first term in (\ref{end-result-cubic}), completing the connection between the inner product of this term and the conservation laws.

\end{document}